%% file: RJwrapper.tex
\documentclass[a4paper]{report}
\usepackage[utf8]{inputenc}
\usepackage[T1]{fontenc}
\usepackage{RJournal}
\usepackage{amsmath,amssymb,array}
\usepackage{booktabs}

\usepackage{subfig}
\usepackage{caption}
\usepackage{float}
\usepackage{multirow}
\usepackage{makecell}

\def\C{{\cal C}}

\def\X{{\cal X}}

\def\Y{{\cal Y}}

\def\log{\hbox{log}}

\def\Beta{\hbox{Beta}}

\def\Dir{\hbox{Dir}}

\def\Ga{\hbox{Ga}}

\def\Mult{\hbox{Mult}}

\def\bse{\begin{eqnarray*}}
\def\ese{\end{eqnarray*}}
\def\be{\begin{eqnarray}}
\def\ee{\end{eqnarray}}
\def\bq{\begin{equation}}
\def\eq{\end{equation}}

\def\th{^{th}}

\def\b1e{{\mathbf e}}
\def\b1f{{\mathbf f}}

\def\bP{{\mathbf P}}

\newcommand{\bmu}{\mbox{\boldmath $\mu$}}

\newcommand{\blambda}{\mbox{\boldmath $\lambda$}}

\begin{document}

%% do not edit, for illustration only
\sectionhead{Contributed research article}
\volume{XX}
\volnumber{YY}
\year{20ZZ}
\month{AAAA}

%% replace RJtemplate with your article
\begin{article}
  \input{BMRMM_Package}

\end{article}

\end{document}

%% file: BMRMM_Package.tex
% !TeX root = RJwrapper.tex
\title{BMRMM: An R Package for Bayesian Markov (Renewal) Mixed Models}
\author{by Yutong Wu and Abhra Sarkar}

\maketitle

% abstract is at most 150-word-long
\abstract{
  We introduce the BMRMM package implementing Bayesian inference for a class of Markov renewal mixed models 
  which can characterize the stochastic dynamics of   
  a collection of sequences, 
  each comprising alternative instances of categorical states and associated continuous duration times, 
  while being influenced by a set of exogenous factors as well as a `random' individual.  
  The default setting flexibly models the state transition probabilities using mixtures of Dirichlet distributions and the duration times using mixtures of gamma kernels 
  while also allowing variable selection for both. 
  Modeling such data using simpler Markov mixed models also remains an option, 
  either by ignoring the duration times altogether  
  or by replacing them with instances of an additional category obtained by discretizing them by a user-specified unit. %above a pre-specified threshold. 
  The option is also useful when data on duration times may not be available in the first place. 
  We demonstrate the package's utility {using two data sets}.
}

\section{Introduction} \label{sec:intro}

Markov models (MMs) are widely used for modeling the transition dynamics of categorical state sequences. 
Classical Markov renewal models (MRMs) additionally model the state duration times, when available, 
where the state transitions follow Markov dynamics 
and the state duration times follow a continuous distribution that depends on the immediately preceding and following states (Figure \ref{fig:graph model}). 
M(R)Ms have been widely used in different variations 
\citep{phelan1990bayes,eichelsbacher2002bayesian,muliere2003reinforced,alvarez2005estimation,
diaconis2006bayesian,bulla2007bayesian,etterson2007analysis,bacallado2009bayesian,
li2009markov,epifani2014bayesian,siebert2016bayesian,holsclaw2017bayesian,sesia2019gene}. 
There are also some sparse works on covariate-dependent Markov models \citep{muenz1985markov,gradner1990analyzing,alioum1998effect,islam2006higher}.

The existing literature however focuses very heavily on modeling single sequences. 
\cite{sarkar2018bayesian} developed a highly flexible and computationally efficient class of Bayesian Markov mixed models (BMMMs) for 
jointly modeling a collection of categorical sequences, 
each one associated with an individual as well as a set of time-invariant external covariates (e.g., the sex and genotype of the associated individual, an experimental condition under which the sequence was generated, etc.). 
BMMMs characterize the state transition probabilities using a convex combination of a fixed covariate-dependent component and a random individual-level component, 
both being Dirichlet distributed. 
They further allow covariate levels with similar effects to be probabilistically clustered together, allowing automatic selection of the significant covariates, and providing a sophisticated framework for analyzing data sets having the aforementioned structure. 

{BMMMs however do not model duration times of the states which are often additionally available in real-world applications.  
Recently, \cite{wu2021bayesian} extended BMMMs to Bayesian Markov renewal mixed models (BMRMMs), 
allowing for the additional analysis of continuous duration times which, depending on the application, can either be the duration for which a state persists or the duration between two consecutive states, i.e., inter-state intervals (ISIs).
Specifically, they modeled the duration times using mixtures of gamma kernels with mixture probabilities being a convex combination a covariate-dependent effect and an individual-level effect, similar to BMMMs. 
Covariate levels with similar influences on mixture probabilities are clustered together in BMRMMs as well, allowing the selection of the significant covariates.  
BMRMMs thus holistically model both state transitions and continuous duration times, painting a comprehensive picture of the underlying stochastic dynamics. %for such data sets. 
}

In this article, we describe the R package \CRANpkg{BMRMM} which implements BMMMs and BMRMMs, collectively referred to henceforth as BM(R)MMs. 
The \CRANpkg{BMRMM} package runs posterior inference for categorical state transitions and continuous duration times, if available, 
via a Markov chain Monte Carlo (MCMC) algorithm, returning an object containing comprehensive inference results. 
The package also includes a suite of plotting functions to display the results graphically. 
Specifically for continuous duration times, when available, the package provides users with three different options: (i)
ignore the duration times and model the state transitions alone as a BMMM; 
(ii) introduce an additional category by discretizing the continuous duration times by a user-specified unit, 
and analyze the appended state transitions as a BMMM; 
(iii) model the duration as a mixture of gamma kernels using a BMRMM, as proposed in \citet{wu2021bayesian}. 
Additionally,  users can choose to turn off one or both of the fixed covariate effects and the random individual effects. 
Overall, the \CRANpkg{BMRMM} package thus gives users a lot of flexibility in handling their data sets, 
providing inferences for both Bayesian Markov renewal or non-renewal models as needed.

The \CRANpkg{BMRMM} package conveniently includes a synthetic \code{foxp2} data set that describes the laboratory study on the role of the FoxP2 gene implicated in speech deficiencies for adult mice,  which is also the motivating application for the methodology of BMMMs and BMRMMs.  
Mutations in the FoxP2 gene have long been associated with  severe deficits in vocal communication for mammals  \citep{macdermot2005identification}. 
Mice with and without the mutation singing under various "social contexts" have thus been studied in many experiments \citep{Fujita_etal:2008,Castellucci_etal:2016,Gaub_etal:2016,Chabout_etal:2016}. 
{The FoxP2 data set \citep{Chabout_etal:2016}, e.g., comprises the sequences of syllables making up the songs as well as the lengths of inter-syllable intervals (ISIs). 
%,  which can both serve as indicators of potential vocal impairment. 
The data set \code{foxp2} included in the \CRANpkg{BMRMM} package is taken from the simulation study of \cite{wu2021bayesian}. 
It is much shorter than the real FoxP2 data set but closely mimics its other aspects and is used} in this paper to demonstrate how to obtain detailed inferences for both syllable transitions and ISI dynamics for a comprehensive analysis of the vocal repertoire in mice with and without the FoxP2 mutation.

{The utility of the \CRANpkg{BMRMM} package goes well beyond the FoxP2 data set. % which had originally motivated the research on BM(R)MMs in \cite{sarkar2018bayesian} and \cite{wu2021bayesian}. 
As described above, the package is designed for scenarios where 
the data set consists of categorical state sequences associated with an individual as well as a number of observed covariates 
where additional data on continuous duration times may or may not be available.
Data sets with such structures are frequently observed in different areas of scientific research  and can potentially benefit from the   \CRANpkg{BMRMM} package. 
For example,  \citet{islam2006higher} analyzed the transitions of different rainfall  orders  in three districts of Bangladesh under three covariates, wind speed, humidity, and maximum temperature. 
In an education assessment study,  \citet{zhang2019scenario}  recorded sequences of writing states, characterized by keystroke logs, for  257 eighth graders of various genders, races, and socioeconomic statuses. 
%\citet{alioum1998effect} studies the transitions of three stages for 3027 HIV patients with covariates including gender and age.  Similarly, 
\citet{combescure2003assessment} estimated the control states  of 371 asthma patients with different body mass indices (BMIs) and disease severity over a four-year-long study and produced the asthma control data set, which 
we will use as an additional example to demonstrate the utility of our package in this paper.  
For such data sets, the \CRANpkg{BMRMM} package provides a flexible, sophisticated, and principled way to model fixed effects of the covariates and random effects of the individuals in both the state transition dynamics and the distribution of the ISIs. }

{Other computer programs for Markov models with covariates include MARKOV \citep{marshall1995markov} and MKVPCI  \citep{alioum2001mkvpci}.
R packages for analyzing discrete Markov models include  \CRANpkg{markovchain} \citep{Spedicato2017Discrete} and \CRANpkg{msm} \citep{jackson2011multi}.}
\CRANpkg{SemiMarkov} \citep{listwon2015semimarkov}  and \CRANpkg{SMM} \citep{barbu2018smm} provide functions for the simulation and estimation of traditional semi-Markov models.   
Some R packages provide the inference of hidden semi-Markov models,   such as  \CRANpkg{mhsmm}  \citep{o2011hidden}  and \CRANpkg{hhsmm} \citep{amini2022hhsmm}. 
 {\citet{ferguson2012mssurv} built the \pkg{msSurv} package  which  provides a nonparametric estimation of semi-Markov models but does not consider covariates.} 
 There are also R packages implementing MRPs in specific application areas.
 For example, \citet{kharrat2019flexible}  introduced the \CRANpkg{Countr} package for flexible regression models based on MRPs. 
 \citet{pustejovsky2021} developed the \CRANpkg{ARPobservation} for simulating behavior streams based on alternating renewal processes.  
 Other R packages for categorical data analysis include \CRANpkg{catdap} \citep{katsura1980catdap} and \CRANpkg{vcd}  \citep{meyer2022vcd}.
 {To our knowledge, there has not been an R package that implements flexible Bayesian M(R)MMs.}

%We summarize the technical details of BMRMMs in Section~\ref{sec:models}.  In Section~\ref{sec:func}, we document the main functions of the \CRANpkg{BMRMM} package.  We then demonstrate our package in analyzing the FoxP2 data set and the asthma control data set in Section~\ref{sec:foxp2} and \ref{sec:asthma}, respectively, and conclude in Section~\ref{sec:end}. 

{In the following section, we summarize the technical details of BMRMMs. 
Documentation for the functions of the  \CRANpkg{BMRMM} package is then provided. 
Next, we demonstrate the usage of our package in analyzing two different data sets. 
The final section contains some concluding remarks.
}

\vspace*{-6pt}
\section{The Bayesian Markov (renewal) mixed models} \label{sec:models}
\begin{figure}[ht!]
    \centering
\includegraphics[width=\textwidth]{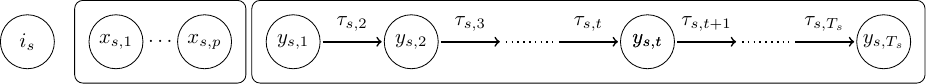}
    \caption{Graphical model showing the data structure: $y_{s,t}$ denotes the observed state at the $t\th$ time location in the $s\th$ sequence; 
    $\tau_{s,t}$ denotes the observed {duration times (either state persistence times or inter-state intervals)} between the states $y_{s,t-1}$ and $y_{s,t}$; 
    each sequence $s$ is also associated with an individual $i_{s}$ and a set of exogenous time-invariant covariates $x_{s,1},\dots, x_{s,p}$. 
    The Markov mixed model considered in this article analyzes the state transitions $y_{s,t}$ in a collection of sequences; 
    the Markov renewal mixed model additionally analyzes the {duration times} $\tau_{s,t}$; 
    both models accommodate fixed effects of the covariates $x_{s,1},\dots, x_{s,p}$ and random effects of the individuals $i_{s}$.
    }
    \label{fig:graph model}
\end{figure}

\vspace*{-4pt}
We briefly describe the BM(R)MM methodologies here -- more details can be found in \citet{sarkar2018bayesian} and \citet{wu2021bayesian}. 
Consider specifically a sequence $s$ comprising $T_s$ state instances and let $y_{s,t}$ denote the state at time $t$ in sequence $s$. 
The states $y_{s,t}$'s come from a set $\Y = \{1,2,\dots,d_0\}$.
Within a sequence $s$, there are $T_s-1$ {duration times (state persistence times or inter-state intervals)}, 
denoted by $\{\tau_{s,t}\}_{s=1,t=2}^{s_0,T_s}$, where $\tau_{s,t}$ is the {duration} between the $(t-1)\th$ and $t\th$ states in sequence $s$, and $s_0$ represents the total number of sequences.
Figure~\ref{fig:graph model} presents a graphical summary of the data structure. 
Each sequence $s$ is associated with $p$ categorical covariates or factors, denoted by $x_{s,j} \in \X_j = \{1,2,\dots,d_j\}$, and an individual, denoted by $i_{s}$. 
Without loss of generality, we assume that the analyses of the transition probabilities and the {duration times} distributions both include all $p$ covariates. 
Moreover, the analysis of {duration times} counts the previous state $y_{s,t-1}$ as an additional $(p+1)\th$ covariate. 
In the \CRANpkg{BMRMM} package, users have the flexibility to select particular covariates for each analysis and exclude the previous state from the analysis of duration times.  
In their original definition in \cite{pyke1961markov}, the duration time $\tau_{s,t}$ in an MRP was allowed to depend on both the preceding state $y_{s,t-1}$ and the following state $y_{s,t}$.
To keep the notation simple and the methodology easy to understand for a broad audience, we however only include the preceding state $y_{s,t-1}$ as a predictor of $\tau_{s,t}$ in this paper.
This analysis can be easily modified to  have the pair $(y_{s,t-1}, y_{s,t})$ as a predictor instead of just $y_{s,t-1}$, as was actually done in \cite{wu2021bayesian}.

\vspace{1em}

\vspace*{-2ex}
\subsection{Model for state transitions} \label{sec: BMMM}

{For a sequence $s$ associated with individual $i$ and covariate levels $x_1,\dots,x_p$, the transition probabilities 
$ \Pr(y_{s,t}=y_{t} \mid  i_{s}=i, x_{s,1}=x_{1}, \dots,x_{s,p}=x_{p}, y_{s,t-1}=y_{t-1})=P_{trans,x_{1},\dots, x_{p}}^{(i)}(y_{t} \mid y_{t-1})$ 
are defined as a convex combination of a fixed covariate effect component $\blambda_{trans,x_1,\dots,x_p}(\cdot\mid y_{t-1})$ and a random effect component $\blambda_{trans}^{(i)}$:}
\be
& \hskip -2ex P_{trans,x_{1},\dots, x_{p}}^{(i)} (y_{t} \mid y_{t-1}) = \pi_{trans,0}^{(i)}(y_{t-1}) \lambda_{trans,x_{1},\dots, x_{p}}(y_{t} \mid y_{t-1}) + \pi_{trans,1}^{(i)}(y_{t-1}) \lambda^{(i)}_{trans}(y_{t} \mid y_{t-1}). ~ \label{eq: mixed Markov model}
\ee 
The coefficients of the convex combination, namely, $\{\pi_{trans,0}^{(i)}(y_{t-1}),\pi_{trans,1}^{(i)}(y_{t-1})\}$ are individual-specific and satisfy $\pi_{trans,1}^{(i)}(y_{t-1})=1-\pi_{trans,0}^{(i)}(y_{t-1})$.

For each covariate $j=1,\dots,p$, 
{it is possible that some covariate levels exert a similar effect on the transition dynamics. 
{For example, the components $\blambda_{trans,x_1=1,x_2,\dots,x_p}(\cdot\mid y_{t-1})$ and $\blambda_{trans,x_1=2,x_2,\dots,x_p}(\cdot\mid y_{t-1})$ would be equal if levels 1 and 2 of covariate 1 have similar influences on transition dynamics for fixed levels for covariates $2, \dots, p$.} 
A clustering mechanism for covariate levels allows the fixed component $\blambda_{trans,x_1,\dots,x_p}(\cdot\mid y_{t-1})$ to be the same for all levels with a similar influence.}
In particular, for covariate $j$, we construct the partition $\C_{trans}^{(j)}=\{\C_{trans,h_{j}}^{(j)}\}_{h_{j}=1}^{k_{trans,j}}$ of its levels, where $k_{trans,j}$ is the number of clusters for covariate $j$ and $h_j$ represents the cluster index.
We introduce latent variables $\{z_{trans,j,\ell}\}_{j=1,\ell=1}^{p,d_{j}}$ that indicate the cluster index for the $\ell\th$ label of covariate $j$.
{Two levels of the covariate $j$, $\ell_1, \ell_2 \in \X_j = \{1,\dots,d_j\}$, are clustered together if and only if $z_{trans,j,\ell_1} = z_{trans,j,\ell_2}$.} 
{For the fixed effects, we then replace the covariate levels $x_1,\dots,x_p$'s with cluster indices $h_1,\dots,h_p$'s and   present the fixed effect as  $\blambda_{trans,h_{1},\dots,h_{p}}(\cdot\mid y_{t-1})$. }

{We set Dirichlet priors for both fixed and individual effect components and let them center around the same mean vector $\blambda_{trans,0}$ to facilitate posterior computation. 
The probability vector $\blambda_{trans,0}$ is also given a Dirichlet prior with mean $\blambda_{trans,00}$ to capture the natural preferences of certain states in $\Y$:  }
\bse
&& \blambda_{trans,h_{1},\dots,h_{p}}(\cdot\mid y_{t-1}) \sim \Dir\left\{\alpha_{trans,0}\lambda_{trans,0}(1 \mid y_{t-1}),\dots,\alpha_{trans,0}\lambda_{trans,0}(d_0 \mid y_{t-1})\right\}, \nonumber \\
&& \blambda^{(i)}_{trans}(\cdot \mid y_{t-1}) \sim \Dir\left\{\alpha^{(0)}_{trans}\lambda_{trans,0}(1 \mid y_{t-1}),\dots,\alpha^{(0)}_{trans}\lambda_{trans,0}(d_0 \mid y_{t-1})\right\}, \nonumber \\
&& \blambda_{trans,0}(\cdot\mid y_{t-1}) \sim \Dir\left\{\alpha_{trans,00}\lambda_{trans,00}(1),\dots,\alpha_{trans,00}\lambda_{trans,00}(d_0)\right\}. \nonumber 
\ese

We present the complete Bayesian hierarchical model for the transition dynamics as
\be
&& (y_{s,t} \mid y_{s,t-1}=y_{t-1}, i_{s}=i, z_{trans,1,x_{s,1}} =h_{1},\dots,z_{trans,p,x_{s,p}} =h_{p})  \sim \nonumber\\
&& \hspace{1cm} \Mult\left\{P_{trans,h_{1},\dots,h_{p}}^{(i)}(1\mid y_{t-1}),\dots,P_{trans,h_{1},\dots,h_{p}}^{(i)}(d_0\mid y_{t-1})\right\}, ~~\text{where} \nonumber\\
&& \bP_{trans,h_{1},\dots,h_{p}}^{(i)} (\cdot\mid y_{t-1}) = \pi_{trans,0}^{(i)}(y_{t-1}) \blambda_{trans,h_{1},\dots,h_{p}}(\cdot\mid y_{t-1}) + \pi_{trans,1}^{(i)}(y_{t-1}) \blambda^{(i)}_{trans}(\cdot\mid y_{t-1}),\nonumber\\
&& z_{trans,j,\ell} \sim \Mult\left\{\mu_{trans,j}(1),\dots,\mu_{trans,j}(d_{j})\right\}, ~~~~~ \bmu_{trans,j} \sim \Dir(\alpha_{trans,j},\dots,\alpha_{trans,j}), \nonumber\\
&& \blambda_{trans,h_{1},\dots,h_{p}}(\cdot\mid y_{t-1}) \sim \Dir\left\{\alpha_{trans,0}\lambda_{trans,0}(1 \mid y_{t-1}),\dots,\alpha_{trans,0}\lambda_{trans,0}(d_0 \mid y_{t-1})\right\},\nonumber \\
&& \blambda_{trans}^{(i)}(\cdot \mid y_{t-1}) \sim \Dir\left\{\alpha^{(0)}_{trans}\lambda_{trans,0}(1 \mid y_{t-1}),\dots,\alpha^{(0)}_{trans}\lambda_{trans,0}(d_0 \mid y_{t-1})\right\}, \nonumber\\
&& \blambda_{trans,0}(\cdot\mid y_{t-1}) \sim \Dir\left\{\alpha_{trans,00}\lambda_{trans,00}(1),\dots,\alpha_{trans,00}\lambda_{trans,00}(d_0)\right\}, \nonumber \\
&& \pi_{trans,0}^{(i)}(y_{t-1}) \sim \Beta(a_{trans,0},a_{trans,1}), \nonumber \\
&& \alpha_{trans,0}\sim\Ga(a_{trans,0},b_{trans,0}),~~~~~~~\alpha^{(0)}_{trans}\sim\Ga(a^{(0)}_{trans},b^{(0)}_{trans}).\nonumber
\ee

\subsection{Model for continuous duration times}
\label{sec:isi}

The \CRANpkg{BMRMM} package provides three options for analyzing the duration times: 
(i) ignore the durations altogether and only model the transition probabilities of the existing states, 
(ii) treat the durations as blocks of a new special category, with a discretization unit specified by users, 
(iii) model the durations as a continuous random variable with a flexible mixture of gamma distributions.
For the first two options, we only need to apply the model described in the previous subsection. 
For the third option, we need to conduct a separate analysis of the duration times as described below.

Let $K$ denote the number of gamma mixture components in the model for the {duration times}.  
{We let $\bP_{dur}^{(i)}(\cdot \mid x_1,\dots,x_p,y_{s,t-1})$ denote the mixture probability vector given individual $i$, covariate levels $x_1,\dots,x_p$,  and preceding syllable $y_{s,t-1}$. 
The preceding state $y_{s,t-1}$ can be removed from the formula if the users do not wish to consider its influence in the inference of the {duration times}. 
The distribution of the continuous {duration times}, $\{\tau_{s,t}\}_{s=1,t=2}^{s_0,T_s}$, is then modeled as }
\vspace{-3ex}\\
\be
& f(\tau_{s,t} \mid  i_{s}=i, x_{s,1}=x_{1}, \dots,x_{s,p}=x_{p}, y_{s,t-1}=y_{t-1}) \nonumber \\
& = \sum_{k=1}^{K}P_{dur}^{(i)}(k \mid x_{1},\dots,x_{p},y_{t-1}) \Ga(\tau_{s,t} \mid \alpha_{k},\beta_{k}),  \nonumber
\ee
{where $\alpha_k$ and $\beta_k$ denote the shape and rate parameters of the $k\th$ gamma mixture component, respectively.  
We  introduce a set of latent variables $\{z_{dur,s,t}\}_{s=1,t=2}^{s_0,T_s}$ that represents the index of the mixture component. 
If $z_{dur,s,t}$ equals to $k$, then $\tau_{s,t}$ follows $\Ga(\alpha_k,\beta_k)$ distribution, i.e., }
\vspace{-3ex}\\
\be
& f(\tau_{s,t} \mid  z_{dur,s,t}=k) \sim \Ga(\tau_{s,t} \mid \alpha_{k},\beta_{k}), \nonumber\\
& \Pr(z_{dur,s,t}=k \mid  i_{s}=i, x_{s,1}=x_{1}, \dots,x_{s,p}=x_{p}, y_{s,t-1}=y_{t-1})=P_{dur}^{(i)}(k \mid x_{1},\dots,x_{p},y_{t-1}) \nonumber. 
\ee

Similar to the model for the transition probabilities,  
{the mixture probabilities are a convex combination of a fixed population-level effect and a random individual-level effect:}
\vspace{-3ex}\\
\bse
&& \bP^{(i)}_{dur}(\cdot \mid x_{1},\dots,x_{p},y_{t-1})=\pi_{dur,0}^{(i)}(\cdot)\blambda_{dur,x_{1},\dots,x_{p},y_{t-1}}(\cdot)+\pi_{dur,1}^{(i)}(\cdot)\blambda^{(i)}_{dur}(\cdot),
\ese
{where $\blambda_{dur,x_{1},\dots,x_{p},y_{t-1}}(\cdot)$ is the baseline component,  $\blambda^{(i)}_{dur}(\cdot)$ is the random individual effect, and $\{\pi_{dur,0}^{(i)}(k),\pi_{dur,1}^{(i)}(k)\}_{k=1}^K$ are individual-specific coefficients such that $\pi_{dur,1}^{(i)}(k)=1-\pi_{dur,0}^{(i)}(k)$.
Again, for each covariate $r=1,\dots,p,p+1$ (where the $(p+1)\th$ covariate is the preceding state $y_{t-1}$),  }
we construct the partition  $\C_{dur}^{(r)}=\{\C_{dur,g_{r}}^{(r)}\}_{g_{r}=1}^{k_{dur,r}}$ of its levels, where $k_{dur,r}$ is the number of clusters for covariate $r$ and $g_{r}$ represents the cluster index. 
We introduce latent variables $\{z_{dur,r,w}\}_{r=1,w=1}^{p+1,d_{r}}$ that indicate the cluster index for the $w\th$ label of the $r\th$ covariate.
{We now replace the population-level effect $\blambda_{dur,x_{1},\dots,x_{p},y_{t-1}}(\cdot)$ with $\blambda_{dur,g_{1},\dots,g_{p},g_{p+1}}(\cdot)$.}

{The mixture probability vectors are given Dirichlet priors with the mean vector $\blambda_{dur,0}$, which itself centers around a global vector $\blambda_{dur,00}$:}
\vspace{-3ex}\\
\bse
&& \blambda_{dur,g_{1},\dots,g_{p+1}}(\cdot) \sim \Dir\left\{\alpha_{dur,0}\lambda_{dur,0}(1),\dots,\alpha_{dur,0}\lambda_{dur,0}(K)\right\}, \nonumber  \\
&& \blambda^{(i)}_{dur}(\cdot) \sim \Dir\left\{\alpha^{(0)}_{dur}\lambda_{dur,0}(1),\dots,\alpha^{(0)}_{dur}\lambda_{dur,0}(K)\right\}, \nonumber\\
&&\blambda_{dur,0}(\cdot) \sim \Dir\left\{\alpha_{dur,00}\lambda_{dur,00}(1),\dots,\alpha_{dur,00}\lambda_{dur,00}(K)\right\}. \nonumber
\ese

We present 
the complete Bayesian hierarchical model for the {continuous duration times} as
\be
&& (\tau_{s,t} \mid z_{dur,s,t}=k) \sim \Ga(\tau_{s,t} \mid \alpha_{k},\beta_{k}), \nonumber\\
&& (z_{dur,s,t} \mid i_{s}=i, z_{dur,1,x_{s,1}} =g_{1}, \dots, z_{dur,p,x_{s,p}}=g_{p}, z_{dur,p+1,y_{s,t-1}}=g_{p+1})  \sim \nonumber\\
&& \hspace{1cm} \Mult\left\{P_{dur,g_{1},\dots,g_{p+1}}^{(i)}(1),\dots,P_{dur,g_{1},\dots,g_{p+1}}^{(i)}(K)\right\}, ~~\text{where} \nonumber\\
&& P^{(i)}_{dur,g_{1},\dots,g_{p+1}}(k)=\pi_{dur,0}^{(i)}(k)\lambda_{dur,g_{1},\dots,g_{p+1}}(k)+\pi_{dur,1}^{(i)}(k)\lambda^{(i)}_{dur}(k), \nonumber \\
&& \blambda_{dur,g_{1},\dots,g_{p+1}}(\cdot) \sim \Dir\left\{\alpha_{dur,0}\lambda_{dur,0}(1),\dots,\alpha_{dur,0}\lambda_{dur,0}(K)\right\}, ~~~\alpha_{dur,0}\sim\Ga(a_{dur,0},b_{dur,0}), \nonumber\\
&& \blambda_{dur}^{(i)}(\cdot) \sim \Dir\left\{\alpha_{dur}^{(0)}\lambda_{dur,0}(1),\dots,\alpha_{dur}^{(0)}\lambda_{dur,0}(K)\right\}, ~~~\alpha_{dur}^{(0)}\sim\Ga(a_{dur}^{(0)},b_{dur}^{(0)}), \nonumber\\
&&\blambda_{dur,0}(\cdot) \sim \Dir\left\{\alpha_{dur,00}\lambda_{dur,00}(1),\dots,\alpha_{dur,00}\lambda_{dur,00}(K)\right\},\nonumber\\
&& \pi_{dur,0}^{(i)}(k) \sim \Beta(a_{dur,0},a_{dur,1}),\nonumber\\ 
&& \alpha_{k} \sim \Ga(a_{dur,0},b_{dur,0}), ~~~\beta_{k} \sim \Ga(a_{dur,0},b_{dur,0}). \nonumber
\ee

{Inference is based on samples drawn from the posterior using a {Metropolis-Hastings-within-Gibbs MCMC algorithm. 
Most full conditionals are available in closed form and can be directly sampled from. 
A Metropolis-Hastings step is however used for updating the discrete valued cluster configurations.} 
%Otherwise, full conditional posterior distributions are derived and sampled using Gibbs.
There is, however, no conjugate prior for gamma distributions with unknown shape parameters \citep{damsleth1975conjugate}. 
Recently, \citet{miller2019fast} designed a procedure that efficiently approximates the posterior full conditionals of gamma shape parameters under a gamma prior with another gamma density. 
We adopt this approximation in our MCMC algorithm. }

\section{The BMRMM R package}\label{sec:func}

\subsection{Package description}

The \CRANpkg{BMRMM} package is developed to implement Bayesian Markov (renewal) mixed models. 
The main function \code{BMRMM} of the package carries out detailed analyses of the state transitions and their {duration times} (if applicable) as described in the previous section.
{Moreover,  the package includes a number of supplementary functions that use the results of the main function to produce numerical summaries, visualizations, and diagnostics.  
Table~\ref{tab:fn} provides a brief description of all functions. }

%\textit{Data Format}

%\textit{Function Descriptions}

%A summary of the functions of the \CRANpkg{BMRMM} package is presented below.
%
%\begin{itemize}
%    \item \code{BMRMM}: Runs the MCMC algorithm for posterior inference of the transition dynamics and also the {duration times} when they are analyzed as a continuous variable.
%    \item \code{summary.BMRMM}: Summarizes the inference results of the BMRMM fit. 
%    \item \code{plot.BMRMMsummary}: Visualizes the BMRMM summary with barplots and heatmaps.  
%    \item \code{hist.BMRMM}: Plots the histogram of the {duration times} superimposed with the posterior mean of the mixture gamma distribution when the {duration times} are analyzed as a continuous variable.  
%    \item \code{diag.BMRMM}: Provides the traceplots and autocorrelation plots for (i) state transitions from the inference of transition probabilities, and/or (ii) the shape and rate parameters by components of the mixture gamma distribution when the {duration times} are analyzed as a continuous variable. 
%    \item \code{model.selection.scores}: Gives the log pseudo marginal likelihood (LPML) \citep{geisser1979predictive} and the widely applicable information criterion (WAIC) \citep{watanabe2010asymptotic} for the number of the components $K$ of the mixture gamma model when the {duration times} are analyzed as a continuous variable.; can be used for selecting the suitable $K$.
%\end{itemize}

\begin{table}[ht!]
    \centering
\resizebox{\columnwidth}{!}{  \begin{tabular}{ll}
    \toprule
       Function  & Description \\\midrule
   \code{BMRMM} & Creates a \code{BMRMM} object.  \\ 
        \code{summary.BMRMM} &  Summary for an object of class  \code{BMRMM} and create a  \code{BMRMMsummary} object. \\
       \code{plot.BMRMMsummary} &  Visualization of  a \code{BMRMMsummary} object.  \\
     \code{hist.BMRMM} & Returns histograms of duration times for a  \code{BMRMM} object.  \\
    \code{diag.BMRMM} &   Provides MCMC diagnostic plots for a  \code{BMRMM} object.  \\
  \code{model.selection.scores} & Returns the LPML and  WAIC scores of the mixture gamma model.\\\bottomrule
    \end{tabular}}
    \caption{Summary of functions in the \CRANpkg{BMRMM} package.}
    \label{tab:fn}
\end{table}

\subsection{The main function BMRMM}

The main function is \code{BMRMM} which implements the inference for both the state transition probabilities and the {duration times}. We summarize the parameters  in Table~\ref{tab:bmrmm} and present the function as follows.

\begin{example}
BMRMM(data, num.cov, cov.labels = NULL, state.labels = NULL, 
      random.effect = TRUE, fixed.effect = TRUE, 
      trans.cov.index = 1:num.cov, duration.cov.index = 1:num.cov, 
      duration.distr = NULL, duration.incl.prev.state = TRUE,
      simsize = 10000, burnin = simsize/2)
\end{example}

{The parameter \code{data} specifies the target data set and  needs to follow a certain structure. } 
The first column should list the individual IDs $i_{s}$, followed by $p$ columns for the values of the $p$ associated covariates $x_{s,j}$, 
then two columns for the values of the previous state $y_{s,t-1}$, the current state $y_{s,t}$, and finally a column for {duration times} $\tau_{s,t}$.
{The package supports {one to five} categorical covariates that take on values ${1,2,\dots}$.} 
The {duration times} column is optional if the user would like to use BMMM instead of BMRMM to analyze just the state transitions.
This is shown in Table~\ref{tab:data-cols}.  
The users can look at the included {simulated data set} \code{foxp2}  as an example.

\begin{table}[h]
    \centering
  \resizebox{\columnwidth}{!}{  \begin{tabular}{ccccccc}
    \toprule
       Id  & Covariate 1 & $\dots$ & Covariate $p$  & Previous State & Current State & {State Durations/ISI}\\\bottomrule
    \end{tabular}}
    \caption{Columns of the desired input data set.}
    \label{tab:data-cols}
\end{table}

{The number of covariates in the data set is specified by the argument \code{num.cov}.
The argument \code{cov.labels} is a list of vectors giving the names of covariate levels in the covariate order that is presented in \code{data} while the parameter \code{state.labels} is a  vector providing the names of the transition states. 
The default labels are Arabic numerals.
} 
{The {\code{random.effect}} parameter gives users the option to exclude the random individual effects. 
If {\code{random.effect}} is set to \code{FALSE}, the transition probabilities (and the mixture probabilities for {duration times}, if applicable) will only consider the influence of the covariate levels.} 
Similarly, the {\code{fixed.effect}} parameter allows users to exclude the fixed population effects. 
{The default values for \code{random.effect} and \code{fixed.effect} are both \code{TRUE}.}
The covariate indices for the two analyses can be specified by setting \code{trans.cov.index} and  \code{duration.cov.index}. 
{We note that indices specified by \code{trans.cov.index} and  \code{duration.cov.index} refer to the index of the covariate when the first covariate is given index 1,  thus different from its index in \code{data}.}

{Users can define \code{duration.distr} in the following three ways.} 

\begin{enumerate}
\item {If users set \code{duration.distr} to be \code{NULL},  which is the default setting,  then the duration times will be ignored and not modeled at all. } 
The BMMM described will be implemented to analyze the existing state transitions alone. 

\item {If \code{duration.distr} is set as \code{list(`mixDirichlet', unit)},  the duration times will be used to construct a new state \code{`dur.state'}, which will be analyzed along with the original set of states. }
The additional argument \code{unit} must be defined and acts both as a threshold and as a block size for {duration times}. 
For example, if the \code{unit} is set to $5$, then for each {duration} value greater than $5$ units, each block of $5$ unit in it will be treated as an instance of a new \code{'dur.state'} state. 
{If there is a state transition from state \code{`a'} to \code{`b'} with a duration time of $15$ seconds and the \code{unit} is specified at $5$ seconds,  then the updated Markov sequence will contain three consecutive \code{`dur.state'} states, i.e.,  \code{(`a',  `dur.state', `dur.state', `dur.state', `b')}.  }
{Since we adopt the floor operation,  a {duration time} of say $17.68$ seconds will also be replaced by three consecutive instances of \code{'dur.state'} states in this example.}  
The BMMM model  will then be implemented to analyze the resulting appended state transitions.

These first two options may naturally result in loss of information and is therefore not recommended when a detailed analysis of the distribution of the {duration times} is warranted. 

\item {If \code{duration.distr} is set to be \code{list(`mixgamma', shape, rate)},  the duration times are modeled as a continuous random variable using a flexible mixture of gamma kernels, as described for a BMRMM model.} 
In this case,  users can specify the prior shape and rate parameters with the \code{shape} and  \code{rate} arguments within the definition of \code{duration.distr}. 
We note that \code{shape} and  \code{rate} must be numeric vectors of the same length. 
\end{enumerate}

By default, we consider the previous state $y_{s,t-1}$ as a covariate when we model the {duration times} as continuous variables, i.e.,  {\code{duration.incl.prev.state}} is set to \code{TRUE}. 
Users can set this parameter to \code{FALSE} if they wish to exclude the previous state when analyzing the {duration times}.
{The remaining parameters \code{simsize} and \code{burnin} denote the total number of MCMC iterations and the number of burn-ins, respectively. }

\begin{table}[ht!]
    \centering
\resizebox{\columnwidth}{!}{\begin{tabular}{lll}
    \toprule
       Argument  & Explanation & Default value \\\midrule
    \code{data} & the data set to be used following the required format & \\ 
         \code{num.cov} &  an integer giving the number of observed covariates in \code{data} & \\
                  {\code{cov.labels}} & a list of vectors giving names of all covariate levels &  {\code{NULL}} \\
                  {\code{state.labels}} &  a vector giving names of the states &  {\code{NULL}} \\
         {\code{random.effect}} &  \code{TRUE} if random individual effects are included &  {\code{TRUE}} \\
                  {\code{fixed.effect}} &  \code{TRUE} if fixed population effects are included &  {\code{TRUE}} \\
         \code{trans.cov.index} & selects the covariates to analyze for transition probabilities & \code{1:num.cov} \\
          \code{duration.cov.index} &   selects the covariates to analyze for duration times &  \code{1:num.cov} \\
                  \code{duration.distr} &   specifies the distribution for duration times & \code{NULL} \\
    {\code{duration.incl.prev.state}} & \code{TRUE} if $y_{t-1}$ acts as a covariate for the analysis of duration times  & {\code{TRUE}}\\
   \code{simsize} &  number of MCMC iterations & 10000\\
   \code{burnin} &  number of burnins of the MCMC algorithm &  \code{simsize}/2\\\bottomrule
    \end{tabular}}
    \caption{Arguments to the \code{BMRMM} function.}
    \label{tab:bmrmm}
\end{table}

The \code{BMRMM} function returns {an object of class \code{BMRMM},  which either contains only  \code{results.trans} or both of  \code{results.trans} and \code{results.duration} if duration times follow a mixture gamma distribution.} 
For the state transitions, the posterior mean transition probability matrices 
for each combination of the covariate levels and each individual are given by \code{results.trans\$tp.exgns.post.mean} and \code{results.trans\$tp.anmls.post.mean}, respectively. 
Additionally, \code{results.trans\$clusters}  stores  cluster configurations for each covariate from each MCMC iteration. 
As for duration times,   the fields \code{results.duration\$shape.samples} and \code{results.duration\$rate.samples} record the  shape  and rate parameters, for each mixture component in every MCMC iteration,  respectively.  
Meanwhile, \code{results.duration\$comp.assignment} gives the assignment of the mixture component for each data point in the last MCMC iteration. 
Similar to transition probabilities, \code{results.duration\$clusters} gives the cluster configurations of the covariates. 
Other elements of \code{results.trans} and \code{results.duration} can be found in the detailed R function description. 

\subsection{{Summarizing BMRMM results}}

{The \CRANpkg{BMRMM} package provides an S3 method for summarizing results of a \code{BMRMM} object as follows. }

\begin{example}
summary.BMRMM(object, delta = 0.02, digits = 2, ...)
\end{example}

{The \code{object} must be of class \code{BMRMM}.  
The argument \code{delta} is associated with local tests for transition probabilities, which we will explain further. 
The \code{digit} parameter is an integer used for number formatting, as in the general \code{summary} function.  
The \code{summary.BMRMM} function returns an object of class \code{BMRMMsummary} with the following fields. }

\begin{itemize}
\item \code{trans.global} and \code{dur.global}

{These two fields give the global test results from the inference of transition probabilities and duration times. 
Global tests show the significance of the covariates in affecting the state transitions and {duration times}. } 
Specifically, for each covariate,  the empirical distribution of the size of the clusters in the stored MCMC iterations is calculated. 
The null hypothesis that a covariate is not important is equivalent to the event that all its levels are in the same cluster, or, in other words, that the cluster size for the covariate is just one.

\item \code{trans.probs.mean} and \code{trans.probs.sd}

The two fields provide the mean and standard deviation for the posterior mean of each transition type under all combinations of covariate levels, respectively. 

\item \code{trans.local.mean.diff} and \code{trans.local.null.test}

{The \code{BMRMMsummary} object also contains local test results for transition probabilities. 
Local tests analyze the differences between the transition probabilities associated with two different levels of a covariate $j$,  fixing the levels of the other covariates.} 
For every pair of levels of covariate $j$,  \code{trans.local.mean.diff} gives the absolute differences in transition probabilities for each transition type in the MCMC iterations.
The local null hypothesis we test for each transition type is that this difference is at least the pre-specified value \code{delta}.
{Meanwhile, \code{trans.local.null.test} gives the probability of the null hypothesis that the difference between two covariate levels is not significant under each transition type. }

\item \code{dur.mix.params} and \code{dur.mix.probs}

For each mixture component,  \code{dur.mix.params} provides the estimates of the gamma shape and rate parameters from the last MCMC iteration.  
{For every covariate level, users can obtain the mixture probabilities by calling the field  \code{dur.mix.probs},  which can be further used to estimate the length of the {duration times}.} 
\end{itemize}

\subsection{{Visualizing results with BMRMM plotting functions}}

{The main plotting function of the package, \code{plot.BMRMMsummary}, is an S3 method  for class \code{BMRMMsummary}.  
It gives the barplots for global tests as well as heatmaps for the posterior mean and standard deviation for transition probabilities, local tests for transition probabilities,  mixture parameters and probabilities for duration times. 
The parameters of \code{plot.BMRMMsummary} include \code{x}, which must be an object of class \code{BMRMMsummary} and \code{type}, which is a single string representing the field of \code{x} that needs to be plotted. 
The function also takes general plotting arguments such as \code{xlab}, \code{ylab}, etc.  } 

\begin{example}
plot.BMRMMsummary(x, type, xlab = NULL, ylab = NULL, main = NULL, col = NULL, ...)
\end{example}

{When duration times are analyzed as continuous variables using mixture gamma distributions, 
the users can use the S3 method  \code{hist.BMRMM} to generate histograms for duration times along with the estimated posterior distribution.
The parameter \code{x} is an object of class \code{BMRMM}. 
The argument \code{comp} gives the specific mixture component that the user would like to investigate. 
When \code{comp} is \code{NULL}, which is the default setting,  the histogram of all observed {duration times} is plotted and superimposed with the posterior mean of the fitted mixture gamma distribution.
When \code{comp} is a specific integer,  we will be looking at the last MCMC iteration. 
The histogram for duration times assigned with component \code{comp} will be presented alongside the mixture gamma distribution with the shape and rate parameters from the last MCMC iteration. 
Users can refer to the documentation of the general \code{hist} function to see the interpretation for the rest of the parameters. 
}

\begin{example}
hist.BMRMM(x, comp = NULL, xlim = NULL, breaks = NULL, main = NULL, 
           col = 'gray', xlab = 'Duration times', ylab = 'Density', ...)
\end{example}

{Finally,  users can check the MCMC diagnostics with the traceplots and autocorrelation plots produced by the function \code{diag.BMRMM}.
The \code{object} parameter should be an object of class \code{BMRMM}. 
For  transition probabilities,  users can specify the covariate levels as well as the state transitions they are interested in by defining \code{cov.combs} and \code{transitions}, respectively. 
For duration times, 
users can define  \code{components}, a numeric vector,  to obtain the diagnostic plots for shape and rate parameters of the specific component kernels.} 

\begin{example}
diag.BMRMM(object, cov.combs = NULL, transitions = NULL, components = NULL) 
\end{example}

\subsection{{Model selection scores for continuous duration times}}
When the {duration times} are modeled using mixtures of gamma distributions, model selection can be performed on the number of mixture components using the function \code{model.selection.scores}.

\begin{example}
model.selection.scores(object) 
\end{example}

{The function takes an \code{object} as its input, which must be an object of class \code{BMRMM}. 
It returns a list consisting of} the log pseudo marginal likelihood (LPML)  \citep{geisser1979predictive} and the widely applicable information criterion (WAIC) \citep{watanabe2010asymptotic} scores of the model.  
Larger values of LPML and smaller values of WAIC indicate better model fits.
They are particularly suitable for complex Bayesian hierarchical models as they can be easily computed from the MCMC samples.

\section{Illustrations on the synthetic FoxP2 data set}\label{sec:foxp2}
The FoxP2 data set records the songs sung by adult male mice of two genotypes, wild type or FoxP2, denoted by $W$ and $F$, respectively \citep{Chabout_etal:2016}. 
The mice sang under three social contexts, $U$ (fresh female urine on a cotton tip placed inside the male's cage), $L$ (an awake and behaving adult female placed inside the cage), and $A$ (an anesthetized female placed on the lid of the cage). 
Each song comprises a sequence of syllables and continuous inter-syllable intervals (ISIs).
The data set can be used to analyze the effect of the FoxP2 gene on the vocal syntax of mice, 
in turn providing insights into the effects of the gene on human vocal communication abilities and related deficiencies.
{The real FoxP2 data set originates from the study by \cite{Chabout_etal:2016} and requires permission to use. 
\cite{wu2021bayesian} simulated a data set that closely mimics the real one.
For demonstrating the \CRANpkg{BMRMM} package, we included in it a shortened version of this synthetic data set 
which we refer to as the \code{foxp2} data set. 
%In the \CRANpkg{BMRMM} package, this shortened synthetic FoxP2 data set is given the name \code{foxp2}.
} 
The \code{foxp2} synthetic data set has $17391$ rows and $6$ columns, which are Id, Genotype, Context, Prev\_State, Cur\_State, and Transformed\_ISI. 
The original FoxP2 data set records ISIs in seconds. 
In the simulated data set \code{foxp2}, following \citet{wu2021bayesian}, log(1+ISI) values are used which give a better model fit. 
%If $\tau_{s,t}$ denotes the original ISI,  the transformed ISI is $\log(1+\tau_{s,t})$. 

\begin{table}[h]
    \centering
    \begin{tabular}{cccccc}
    \toprule
Id & Genotype& Context &Prev\_State& Cur\_State    &    {Transformed\_ISI} \\\midrule
1     &   2     &  2      &    3     &    3 & 0.20197711\\
1     &   2     &  2      &    3     &    3 & 0.06972753\\
1     &   2     &  2      &    3     &    3 & 0.07211320\\
1     &   2     &  2      &    3     &    3 & 0.15790932\\
1     &   2     &  2      &    3     &    3 & 0.06781471\\
1     &   2     &  2      &    3     &    3 & 0.09426236\\\bottomrule
    \end{tabular}
    \caption{Part of the simulated FoxP2 data set \code{foxp2}.}
    \label{tab:foxp2}
\end{table}

If we are only interested in analyzing the transition probabilities with the covariates genotype and social contexts, we would use the main function as follows.

\begin{example}
R> res.fp2 <- BMRMM(foxp2, num.cov = 2)
\end{example}

If we would like to pick specific covariates for our analyses, we can define \texttt{trans.cov.index} and \texttt{duration.cov.index} accordingly. 
For example, if we only want to use context for transition probabilities and genotype for ISIs, we would run the following.

\begin{example}
R> res.fp2 <- BMRMM(foxp2, num.cov = 2, 
                    trans.cov.index = c(2), duration.cov.index = c(1))
\end{example}

If we would like to analyze the ISIs as part of the original state sequence following a mixture Dirichlet distribution, as was done by \citet{sarkar2018bayesian}, 
the ISIs are replaced by (possibly consecutive) "silent" states by dividing them into blocks of 250 milliseconds.
The \code{BMRMM} function can do this by setting \code{duration.distr} as a list with the string \code{'mixDirichlet'} and the argument \code{unit} as $\log(0.25+1)$, based on the log transformation.

\begin{example}
R> res.fp2 <- BMRMM(foxp2, num.cov = 2,  
                    duration.distr = list('mixDirichlet', unit = log(0.25+1)))
\end{example}

In the next example, we would like to analyze the ISIs as continuous variables following a mixture gamma distribution.
For syllable transitions, we use both genotype and context as covariates. 
For the ISIs, in addition to these two, we also use the preceding syllable as a covariate. 

\begin{example}
R> res.fp2 <- BMRMM(data = foxp2, num.cov = 2, state.labels = c('d', 'm', 's', 'u'), 
                    cov.labels = list(c('F', 'W'), c('U', 'L', 'A')),
                    duration.distr = list('mixgamma', shape = rep(1, 4), rate = rep(1, 4)))
\end{example}

In what follows, we show the results for the last function call.
The returned \code{res.fp2} have two parts, which are named  \code{res.fp2\$results.trans} and \code{res.fp2\$results.duration}.
Now we demonstrate how we print and visualize the results.  

\begin{figure}[!ht]
\centering
\subfloat{
\includegraphics[width=0.47\textwidth]{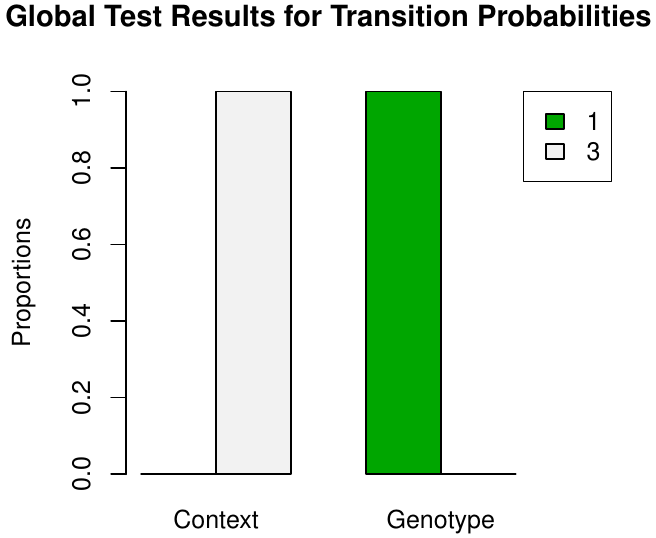}}
\qquad
\subfloat{
\includegraphics[width=0.44\textwidth]{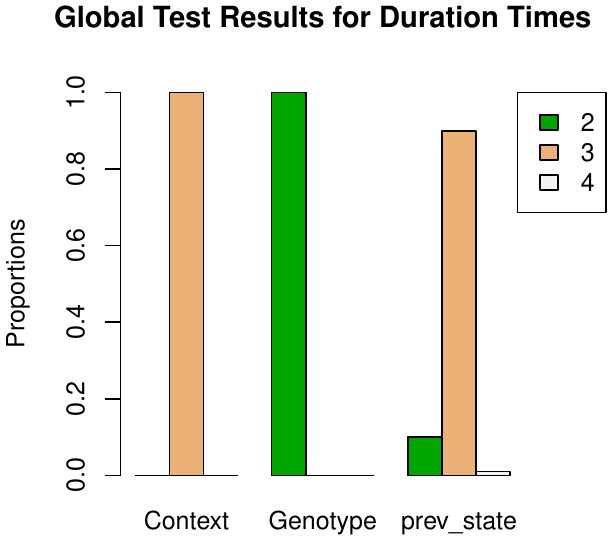}}
\caption{Results for the {simulated \code{foxp2}} data set showing the global tests of significance of the covariates for the state transitions (left) and the ISIs (right).
The bars represent the estimated posterior probabilities of the number of clusters formed by the levels of each covariate.}
\label{fig:global}
\end{figure}

{First, we obtain a \code{BMRMMsummary} object, \code{sm.fp2},  by calling the \code{summary.BMRMM} function on the returned results \code{res.fp2}.
The global test results for  identifying the significant covariates can be found by calling the fields \code{trans.global} and \code{dur.global}.
The function \code{plot.BMRMMsummary} is called to  visualize the global tests using barplots, as presented  in Figure~\ref{fig:global}. } 
{We recall that a covariate is significant when its levels formed more than one cluster with very high posterior probability (the bar heights). 
Figure~\ref{fig:global} and the printed results suggest that every covariate is significant for the ISIs but only the social context is significant for the transition probabilities. }

\begin{example}
R> sm.fp2 <- summary(res.fp2)
R> sm.fp2$trans.global
            label_data
cluster_data Context Genotype
           1       0        1
           3       1        0
R> sm.fp2$dur.global
            label_data
cluster_data Context Genotype prev_state
           2    0.00     1.00       0.10
           3    1.00     0.00       0.90
           4    0.00     0.00       0.01
R> plot(sm.fp2, 'trans.global')
R> plot(sm.fp2, 'dur.global')
\end{example}

\begin{figure}[!ht]
\centering
\subfloat{
\includegraphics[width=.91\textwidth]{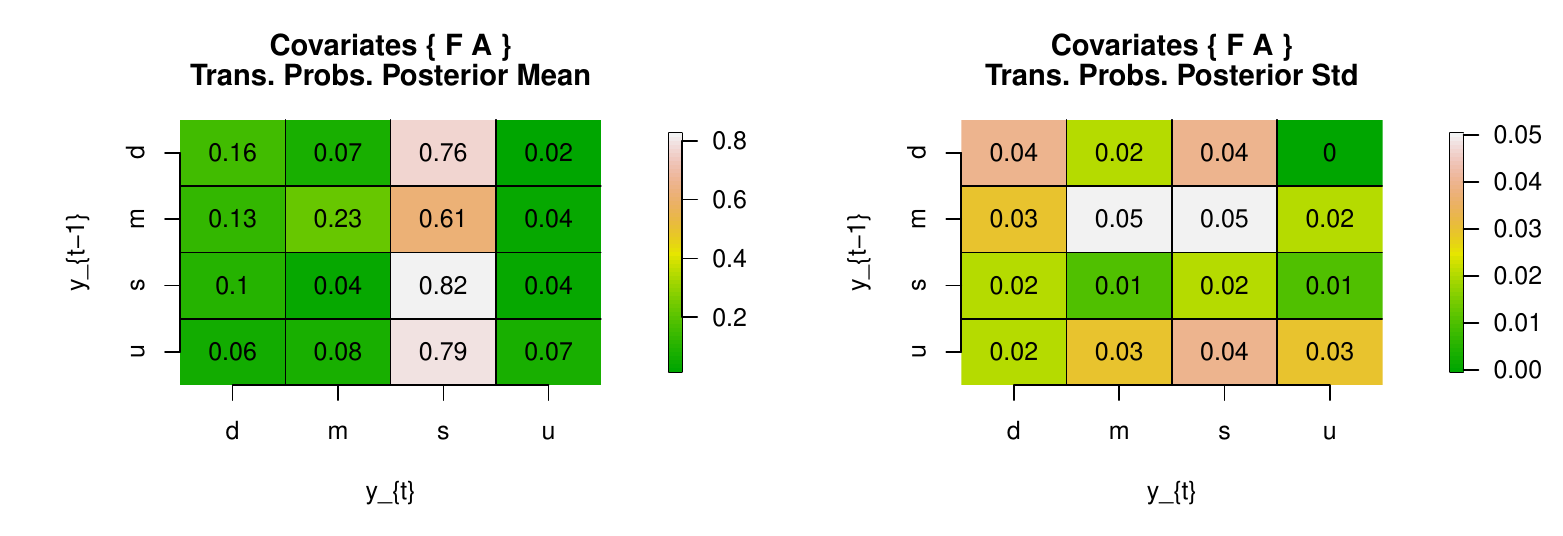}
\label{fig:post_mean_FA}}\\
\subfloat{
\includegraphics[width=.91\textwidth]{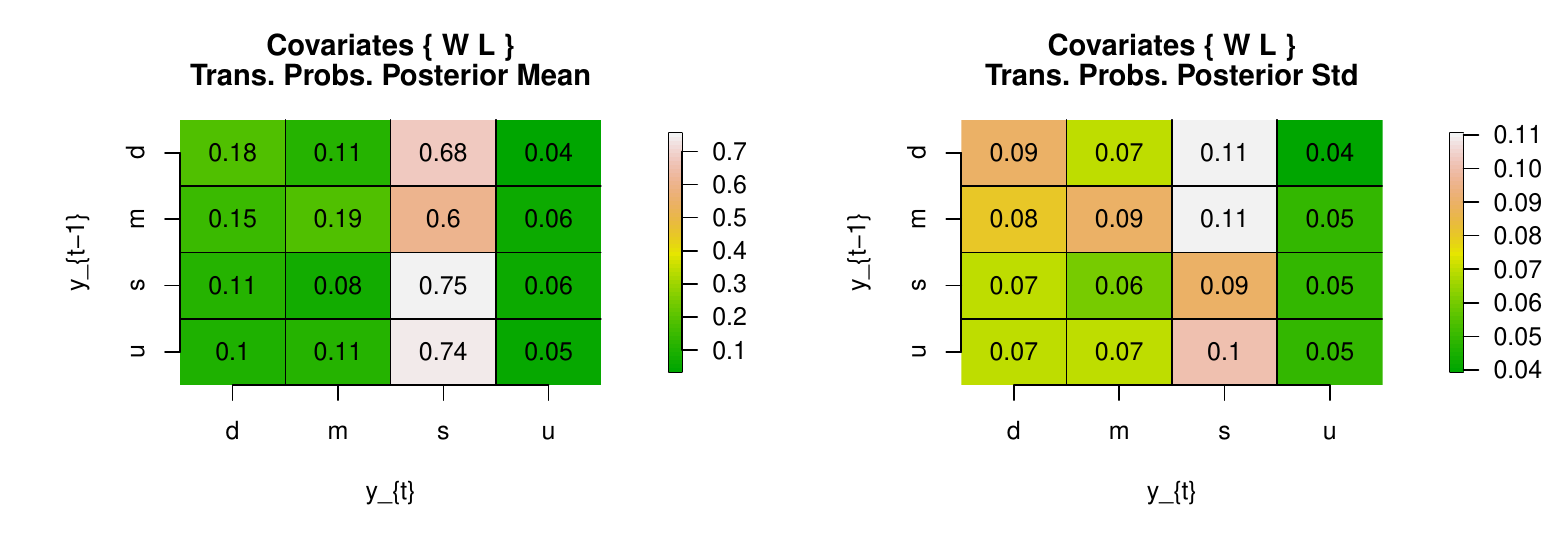}
\label{fig:post_mean_WL}}\\
\caption{Results for the  {simulated \code{foxp2}} data set showing the posterior mean and standard deviation for each transition type for selected covariate combinations, $(F,A)$ (top) and $(W,L)$ (bottom).
 }
\label{fig:post_mean}
\end{figure}

{The plotting function can be called to visualize the posterior transition probabilities under different combinations of the covariate levels. }
We show in Figure~\ref{fig:post_mean} the heatmaps for the posterior mean and standard deviation of the transition probabilities for each transition type for the following combinations of covariates: $(F,A)$ and $(W,L)$.

\begin{example}
R> plot(sm.fp2, 'trans.probs.mean')
R> plot(sm.fp2, 'trans.probs.sd')
\end{example}

\begin{figure}[!ht]
\centering
\subfloat{
\includegraphics[width=.88\textwidth]{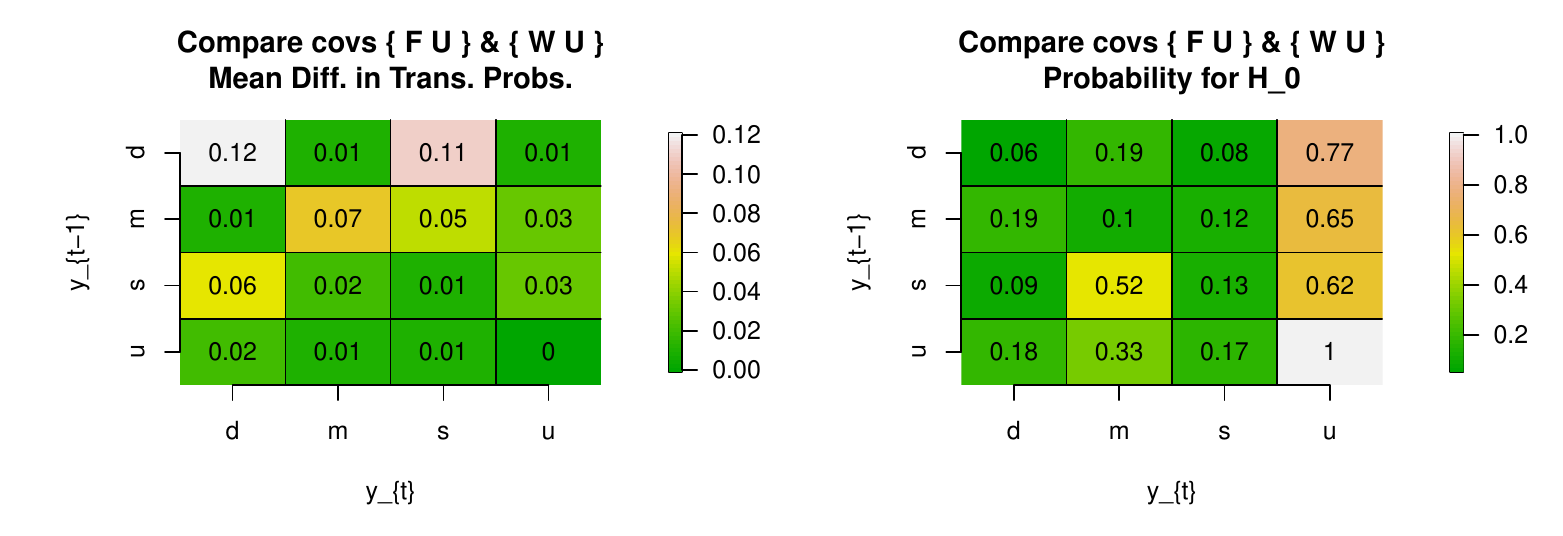}}\\
\subfloat{
\includegraphics[width=.88\textwidth]{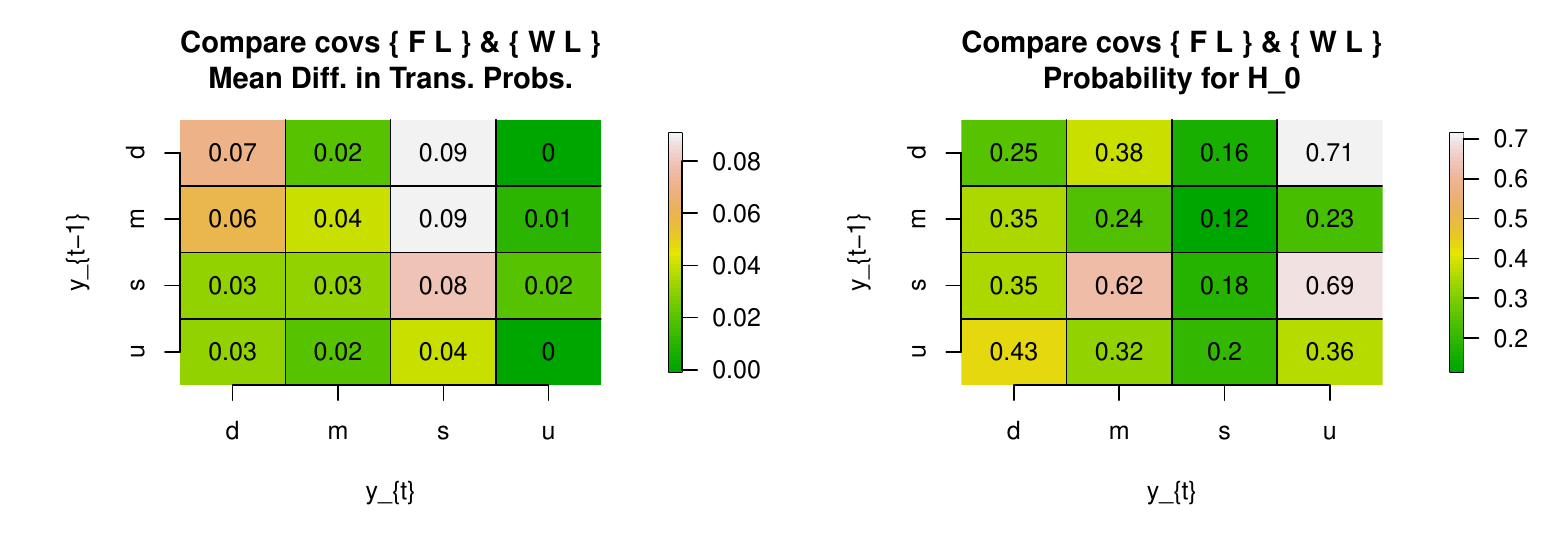}}\\
\subfloat{
\includegraphics[width=.88\textwidth]{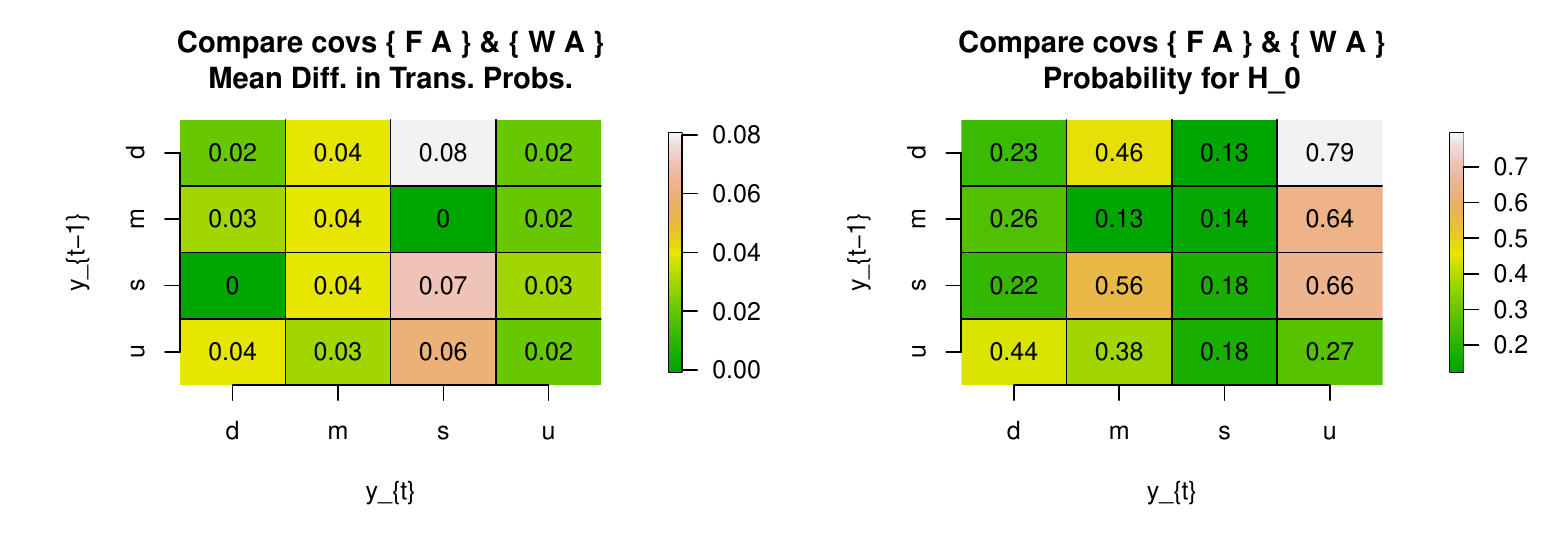}}\\
\caption{Results for the {simulated \code{foxp2}} data set showing local test results for genotypes fixing the social context, $U$ (top), $L$ (middle), and $A$ (bottom).
The averaged absolute difference in transition probabilities between $F$ and $W$ is presented on the left.
The posterior probabilities of the corresponding null hypotheses are on the right.
 }
\label{fig:local}
\end{figure}

We also perform the local test  to assess the influence of genotype on the transition probabilities by computing the absolute difference of the transition probabilities between $F$ and $W$ among the thinned MCMC samples after burn-ins, i.e., $|\Delta \lambda_{trans,\cdot,x_2}(y_t\mid y_{t-1})|=|\lambda_{trans,1,x_2}(y_t\mid y_{t-1})-\lambda_{trans,2,x_2}(y_t\mid y_{t-1})|$. 
The estimated posterior probability for the null hypothesis is therefore the proportion of times $|\Delta \lambda_{trans,\cdot,x_2}(y_t\mid y_{t-1}) \le \delta|$ is observed in the MCMC samples, 
where $x_2$ is the social context and $\delta$ is the user-specific difference threshold \code{delta}. 
{The plotting function \code{plot.BMRMMsummary} gives the plots for all local test results if we set the \code{type} to be \code{`trans.local.mean.diff'} or \code{`trans.local.null.test'}. 
Here, we show the results of local tests for the covariate 1 (i.e., genotype) with \code{delta} equaling the default value of 0.02, and present  the plots in Figure~\ref{fig:local}.}
{From the figure, we see that the posterior probabilities of the null hypotheses are generally large for most transition types (e.g., transitions to the syllable $u$) regardless of the social context, 
indicating that genotype does not have a strong influence on transition probabilities with a fixed context under these transition types. }

\begin{example}
R> plot(sm.fp2, 'trans.local.mean.diff')
R> plot(sm.fp2, 'trans.local.null.test')
\end{example}

\begin{figure}[!ht]
\centering
\subfloat[Histogram of ISIs with the estimated posterior mean (red line) of their marginal gamma mixture density averaged from recorded MCMC samples.]{
\includegraphics[width=0.44\textwidth]{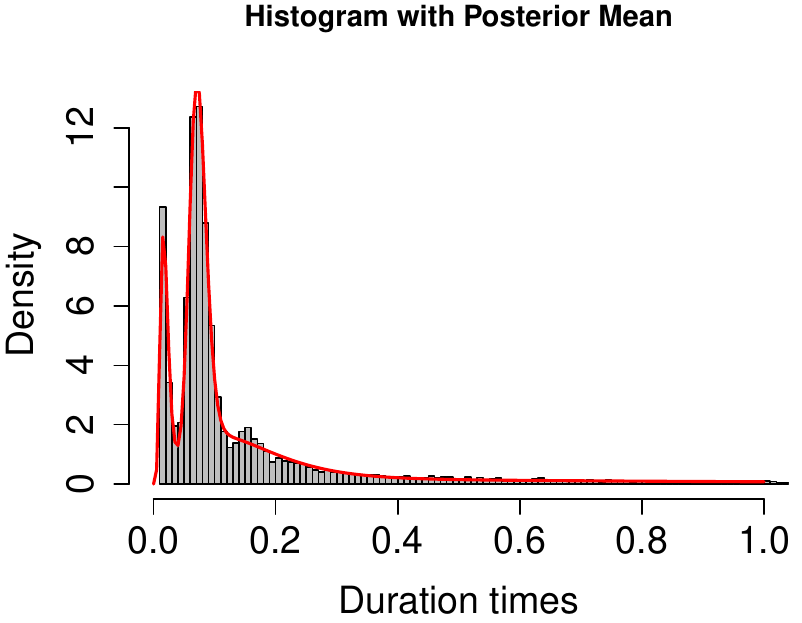}
\label{fig:hist_isi}}
\qquad
\subfloat[Histograms of ISIs for each component of the gamma mixture model along with the component density (red lines) from the last MCMC iteration.]{
\includegraphics[width=0.46\textwidth]{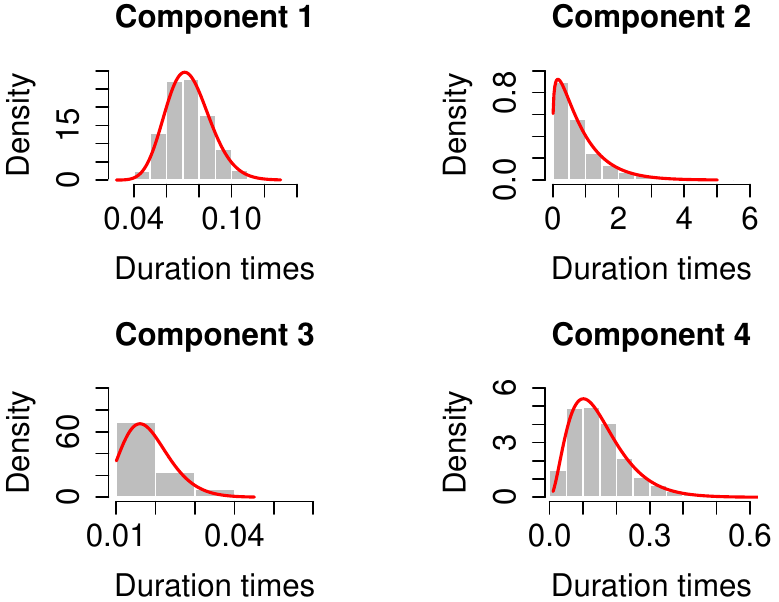}
\label{fig:hist_by_comp}}\\
\caption{Results for the {simulated \code{foxp2}} data set showing the histograms of the ISIs with the estimated posterior gamma mixture density (left) and the histograms of the ISIs for each mixture component (right). 
 }
\label{fig:hist}
\end{figure}

Next, we turn our attention to the ISIs. 
We first check the fit of our estimated mixture gamma distribution presented in Figure~\ref{fig:hist_isi}. 
We then look further into the shape of each mixture component in Figure~\ref{fig:hist_by_comp}. 
{From the histogram for each component, we see that components 2 and 4 represent longer ISIs while components 1 and 3 represent shorter ISIs.}

\begin{example}
R> hist(res.fp2, xlim = c(0,1))
R> for(comp in 1:4) {
     hist(res.fp2, comp = comp)
   }
\end{example}

We examine the values of mixture parameters and mixture probabilities for each covariate level in the last MCMC iteration, which provides insights into the influence of the covariate on ISI lengths.

%Heatmap for Covariate Genotype 
%Heatmap for Covariate Context 
%Heatmap for Covariate prev_state 

\begin{example}                                           
R> sm.fp2$dur.mix.params
       shape.k rate.k
Comp 1   29.07 394.30
Comp 2    1.23   1.49
Comp 3    8.46 465.66
Comp 4    3.03  20.13

R> sm.fp2$dur.mix.probs
$Genotype
          F    W
Comp 1 0.46 0.48
Comp 2 0.19 0.13
Comp 3 0.10 0.15
Comp 4 0.25 0.24

$Context 
          U    L    A
Comp 1 0.58 0.35 0.48
Comp 2 0.14 0.16 0.18
Comp 3 0.08 0.21 0.08
Comp 4 0.20 0.28 0.25

$prev_state 
          d    m    s    u
Comp 1 0.52 0.52 0.42 0.42
Comp 2 0.13 0.13 0.22 0.17
Comp 3 0.11 0.11 0.10 0.18
Comp 4 0.24 0.24 0.26 0.23
\end{example}

{From the mixture probabilities, we see that mice with genotype $F$ have a much higher mixture probability in component 2 compared to genotype $W$, which indicates mice with the FoxP2 mutation require a longer ISI between pronouncing two syllables, a reflection of vocal impairment. }

\begin{figure}[!ht]
\centering
\subfloat[For transition type $u \rightarrow m$ under covariates $\{F,U\}$.]{
\includegraphics[width=.85\textwidth]{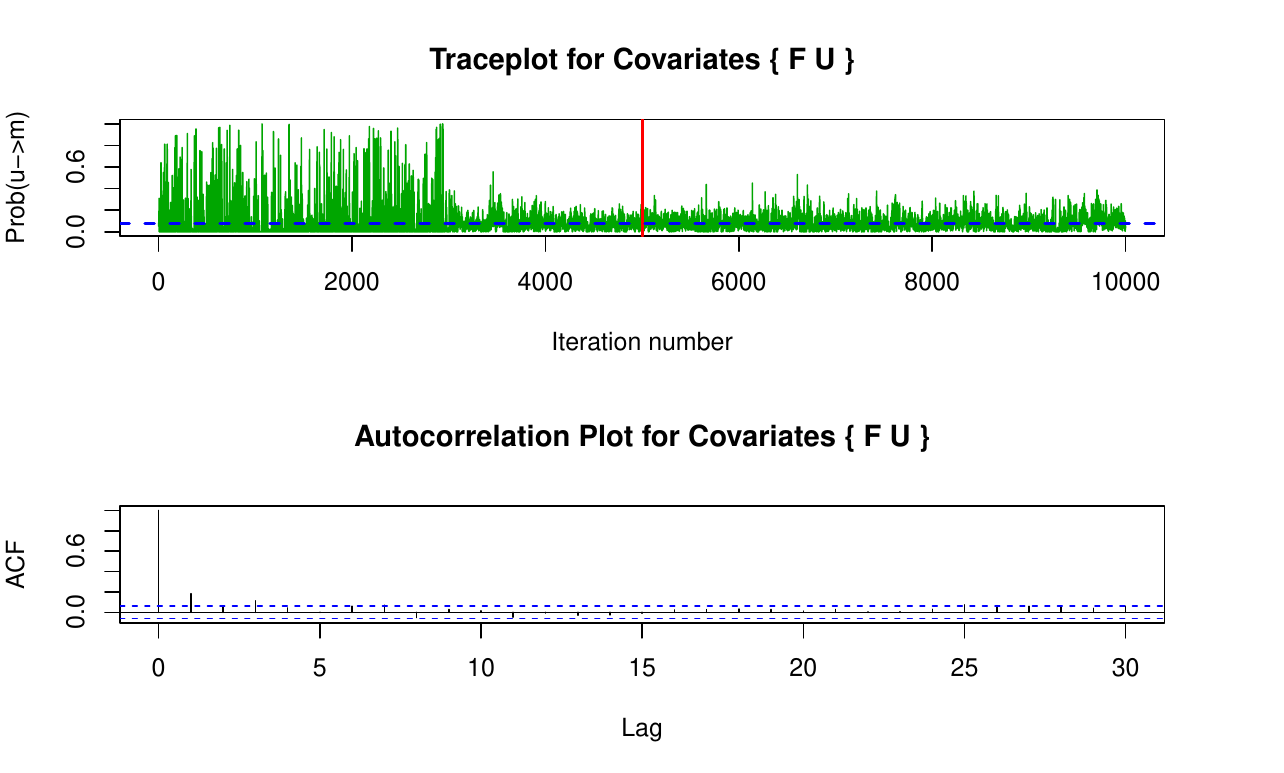}}\\
\subfloat[For gamma mixture component 2 shape and rate parameters.]{
\includegraphics[width=\textwidth]{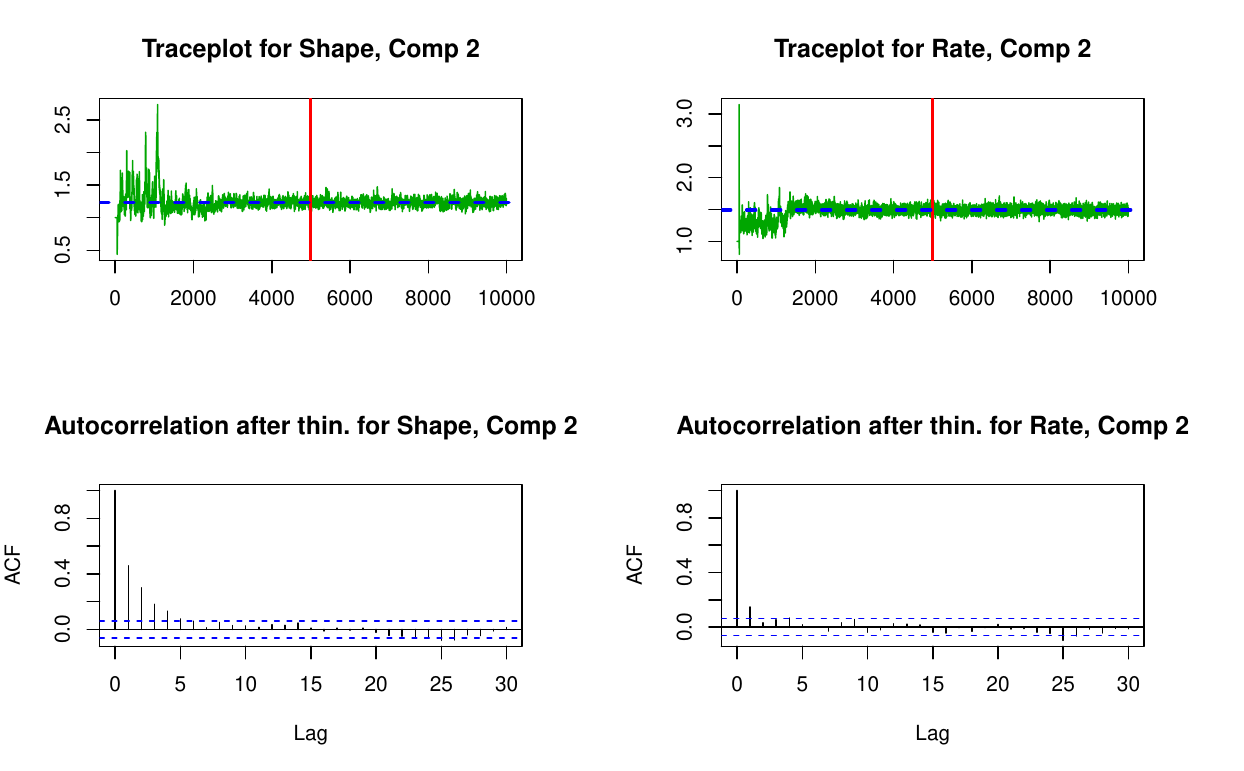}}\\
\caption{Results for the {simulated \code{foxp2}} data set showing the MCMC diagnostic plots, including traceplots and autocorrelation plots. }
\label{fig:diag}
\end{figure}

Finally, we check the MCMC diagnostic plots and see if we had good mixing for the parameters. 
Here we focus on a specific transition type, $u\rightarrow m$, for covariate combination $\{F,U\}$ and a specific mixture component (component 2).
We show these plots in Figure~\ref{fig:diag}.

\begin{example}
R> diag.BMRMM(res.fp2, cov.combs = list(c(1, 1)), 
              transitions = list(c(4, 2)), components = c(2))
\end{example}

\section{Illustrations on the asthma control data set}\label{sec:asthma}
{The \CRANpkg{BMRMM} package is able to analyze duration times in detail which could either be the ISIs, as seen in the synthetic \code{foxp2} data set, or the state persistence times, as in a traditional semi-Markov model. 
To demonstrate the usage of our package in analyzing the state persistence times, we use the asthma control data set from the ARIA (Association pour la Recherche en Intelligence Artificielle) study of  severe asthmatic patients \citep{combescure2003assessment} in France between 1997 and 2001. 
At each visit, a chest physician graded the asthma control status of the patient using control scores \citep{juniper1999development}.  
The data set contains the sojourn time of the control states as well as three  covariates: Asthma severity, sex, and the body mass index (BMI) of the patients. 
\citet{saint2003analysis} used a Markov model with piece-wise constant intensities to model the asthma control evolution and proposed a regression model for analyzing the effect of covariates.
\citet{combescure2003assessment} used the data set to assess the relationship between asthma severity and control of asthma.
\citet{listwon2015semimarkov} fitted a semi-Markov model for the sojourn times using exponential and Weibull distributions and analyzed the effect of covariates individually due to complexity. 
Our \CRANpkg{BMRMM} package is able to analyze the effect of the three covariates while also incorporating random individual effects exhibited by different patients on transition dynamics and state duration times. }

{The \code{asthma} data set we use here is from the \CRANpkg{SemiMarkov} package \citep{listwon2015semimarkov}.
We have renamed and reordered the columns such that the data set fits the required format.}
Specifically, the data set has $928$ rows, recording the asthma control states of $371$ patients, which is one of the following three transient states:  Optimal control (State 1), sub-optimal control (State 2), and unacceptable control (State 3). 
Each state can transit to any other two states and the state duration times are recorded. 
The data set also contains three binary covariates of the asthma patients, including the disease severity (1 if mild-moderate and 2 if severe), BMI (body mass index, 1 if BMI $<25$ and 2 otherwise), and sex (1 if women and 2 if men). 
We display part of the processed data in Table~\ref{tab:asthma}, where \code{Duration} is the sojourn time in \code{Prev\_State}.

\begin{table}[h]
    \centering
    \begin{tabular}{ccccccc}
    \toprule
Id & Severity& BMI & Sex &Prev\_State & Cur\_State    &    Duration \\\midrule
2     &   2     &  2      &  1&  3     &    2 & 0.1533\\
2    &   2     &  2      &    1&2     &    2 & 4.1232\\
3     &   2     &  2      &    2&3     &    1 & 0.0958\\
3     &   2     &  2      &    2&1     &    3 & 0.2300\\
3    &   2     &  2      &    2&3     &    1 & 0.2656\\
3     &   2     &  2      &    2&1     &    1 & 5.4073\\\bottomrule
    \end{tabular}
    \caption{Part of the \code{asthma} data set from the ARIA study of severe asthmatic patients.}
    \label{tab:asthma}
\end{table}

{We investigate the transition dynamics and state persistence times of the \code{asthma} data set using the \code{BMRMM} function. 
We consider $K=4$ mixture components for state persistence times. 
The choice of $K$ is derived from running the BMRMM model several times with different $K$'s and comparing the fitness of the models using the LPML and WAIC scores. }

\begin{example}
R> res.asm <- BMRMM(data = asthma, num.cov = 3, state.labels = c(1, 2, 3), 
                    cov.labels = list(c('Mild-Moderate','Severe'), 
                                      c('BMI<25','BMI>=25'), 
                                      c('Women','Men')),
                    duration.distr = list('mixgamma', shape = rep(1, 4), 
                                                       rate = rep(1, 4)))
\end{example}

\begin{figure}[!ht]
\centering
\subfloat{
\includegraphics[width=0.45\textwidth]{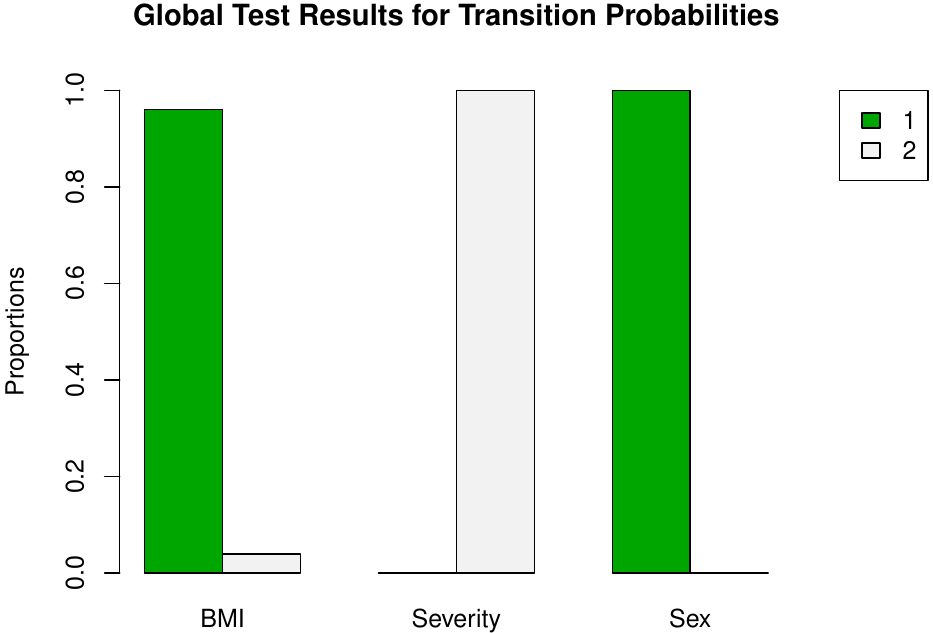}}
\qquad
\subfloat{
\includegraphics[width=0.45\textwidth]{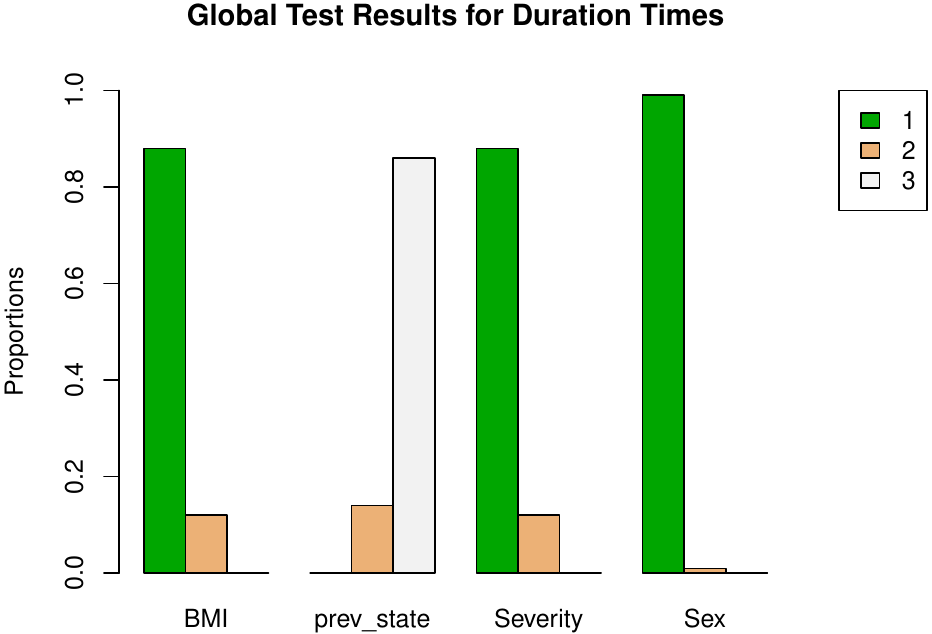}}
\caption{Results for the \code{asthma} data set showing the global tests of significance of the covariates for the state transitions (left) and the state persistence times (right).
The bars represent the estimated posterior probabilities of the number of clusters formed by the levels of each covariate.}
\label{fig:global_asthma}
\end{figure}

{We name the returned \code{BMRMM} object \code{res.asm} and obtain the \code{BMRMMsummary} object \code{sm.asm} by calling the \code{summary.BMRMM} function. 
As in the FoxP2 application, we first plot the global test results for both transition probabilities and state persistence times in Figure~\ref{fig:global_asthma}.
For the transition probabilities, only the severity of asthma is significant while for duration times only the preceding state is significant.
The BMI value and the sex are not significant for either transition dynamics or state durations.}

\begin{example}
R> sm.asm <- summary(res.asm)
R> sm.asm$trans.global
            label_data
cluster_data  BMI Severity  Sex
           1 0.96     0.00 1.00
           2 0.04     1.00 0.00
R> sm.asm$dur.global
                       label_data
cluster_data  BMI prev_state Severity  Sex
           1 0.88       0.00     0.88 0.99
           2 0.12       0.14     0.12 0.01
           3 0.00       0.86     0.00 0.00
R> plot(sm.asm, 'trans.global')
R> plot(sm.asm, 'dur.global')
\end{example}

\begin{figure}[!ht]
\centering
\subfloat{
\includegraphics[width=.92\textwidth]{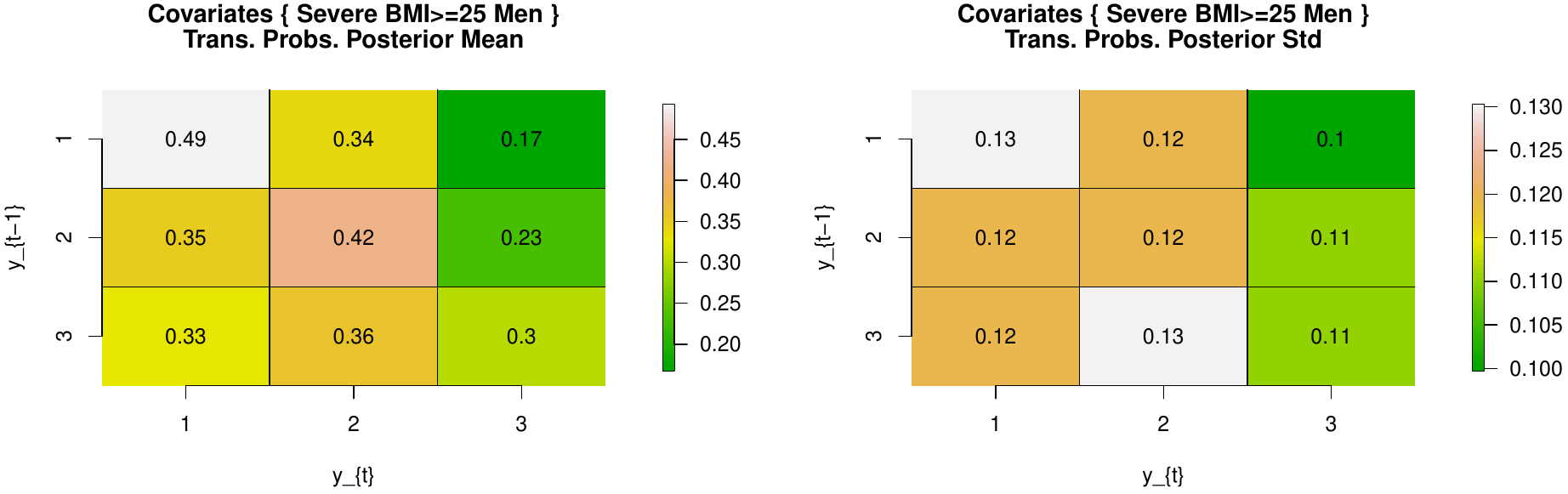}
\label{fig:asthma_sev_men}}\\
\subfloat{
\includegraphics[width=.92\textwidth]{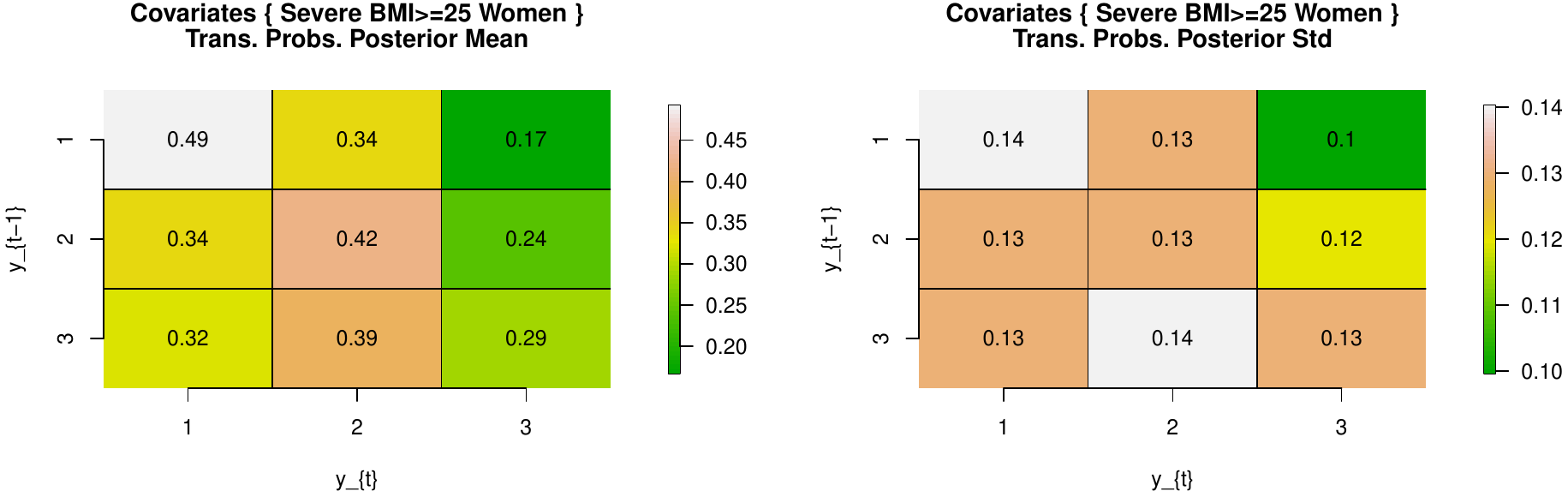}
\label{fig:asthma_sev_women}}\\
\caption{Results for the \code{asthma} data set showing the posterior mean and standard deviation for each transition type for selected covariate combinations, $\{\text{Severe, BMI}\ge25,\text{Men}\}$ (top) and $\{\text{Severe, BMI}\ge25,\text{Women}\}$ (bottom).
 }
\label{fig:post_mean_asthma}
\end{figure}

\begin{figure}[!ht]
\centering
\subfloat{
\includegraphics[width=.92\textwidth]{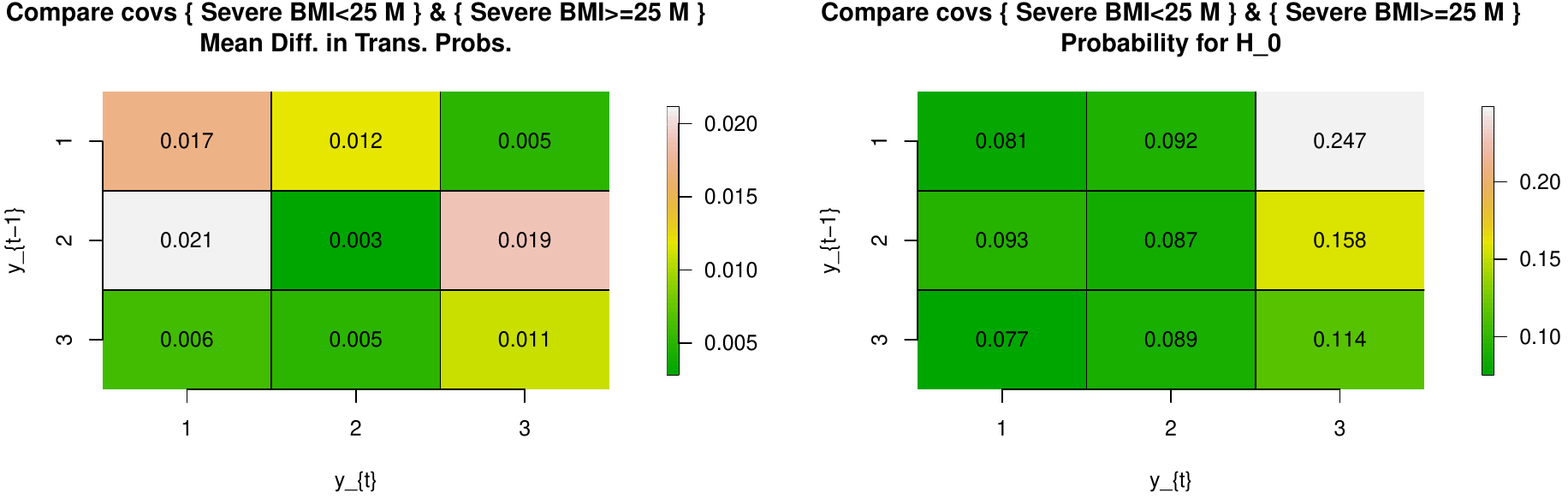}}\\
\subfloat{
\includegraphics[width=.92\textwidth]{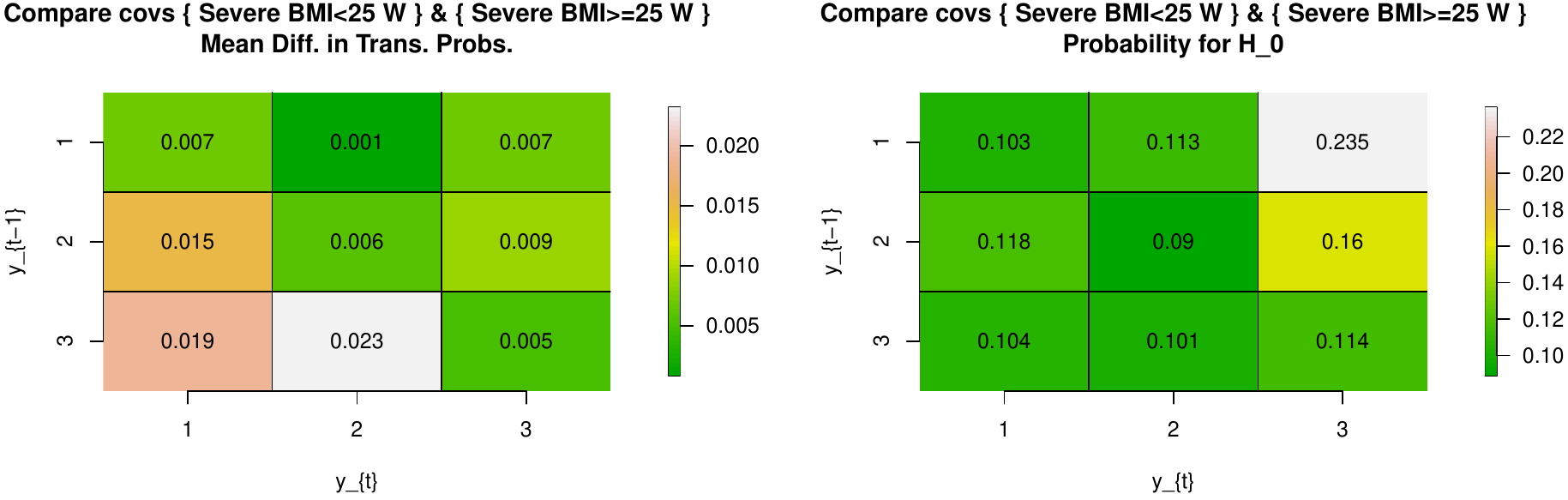}}\\
\caption{Results for the \code{asthma} data set showing local test results for BMI fixing asthma severity condition and sex,  men (top) and women (bottom).
The averaged absolute difference in transition probabilities between BMI $<25$ and BMI $\ge25$ is presented on the left.
The posterior probability of the null hypothesis is on the right.
 }
\label{fig:local_asthma}
\end{figure}

We show the posterior mean and standard deviations of the state transition probabilities 
for men and women with severe conditions and BMI $\ge25$ in Figure~\ref{fig:post_mean_asthma}.
We see that for severe patients with BMI $\ge25$,  the transition probabilities are similar for men and women. 
We also take a look at the local test results for the BMI values fixing the severity of the patients in Figure~\ref{fig:local_asthma}. 
Though the absolute differences between the two covariate levels for BMI are small, the probabilities for the null hypotheses are also small, especially for transitions to state 1 and state 2.   
This suggests that even though the influence of BMI on state transitions is not significant globally,  
it is significant given the severity of asthma condition regardless of sex.

\begin{example}
R> plot(sm.asm, 'trans.probs.mean')
R> plot(sm.asm, 'trans.probs.sd')
\end{example}

\begin{figure}[!ht]
\centering
\subfloat[Histogram of ISIs with the estimated posterior mean (red line) of their marginal gamma mixture density averaged from recorded MCMC samples.]{
\includegraphics[width=0.45\textwidth]{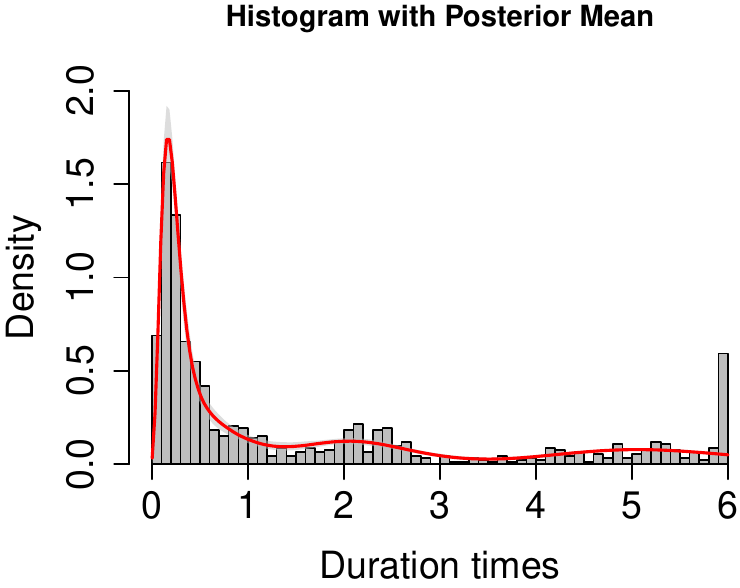}
\label{fig:hist_isi_asthma}}
\qquad
\subfloat[Histograms of ISIs for each of the three components of the gamma mixture model along with the component density (red lines) from the last MCMC iteration.]{
\includegraphics[width=0.45\textwidth]{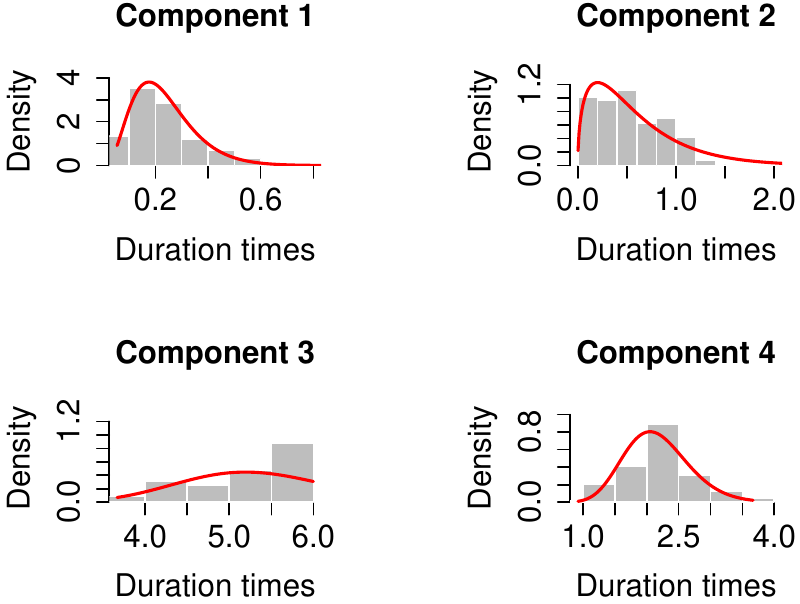}
\label{fig:hist_by_comp_asthma}}\\
\caption{Results for the \code{asthma} data set showing the histograms of ISIs with the estimated posterior gamma mixture density (left) and histograms of ISIs for each mixture component (right). 
 }
\label{fig:hist_asthma}
\end{figure}

{Figure~\ref{fig:hist_isi_asthma} presents the histograms of the entire asthma data set, superimposed with the posterior mean of the mixture gamma distribution. 
With four components, the estimated mixture gamma fits the asthma data well. 
From the histogram for each component, we see from Figure~\ref{fig:hist_by_comp_asthma} that component 1 and 2 represents shorter state persistence times while components 3 and 4 represent longer durations.}

\begin{example}
R> hist(res.asm, xlim = c(0,1))
R> for(comp in 1:4) {
     hist(res.asm, comp = comp)
   }
\end{example}

\begin{figure}[!ht]
\centering
\subfloat[For transition type $1 \rightarrow 1$ under covariates \{Mild-Moderate, BMI $\ge$ 25, Women\}.]{
\includegraphics[width=.85\textwidth]{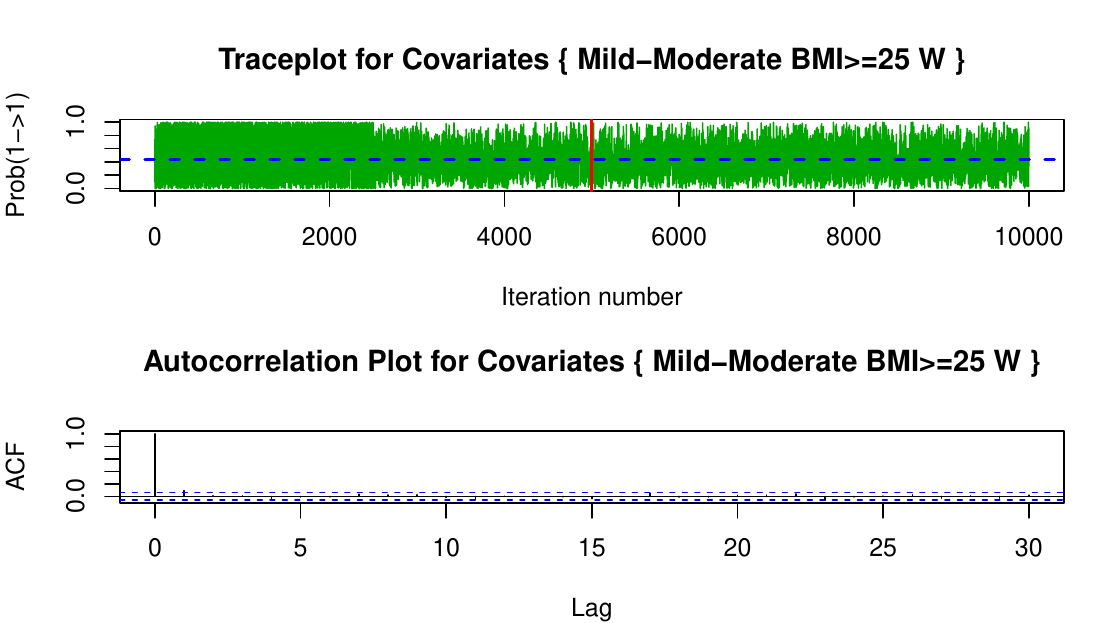}}\\
\subfloat[For gamma mixture component 3 shape and rate parameters.]{
\includegraphics[width=.85\textwidth]{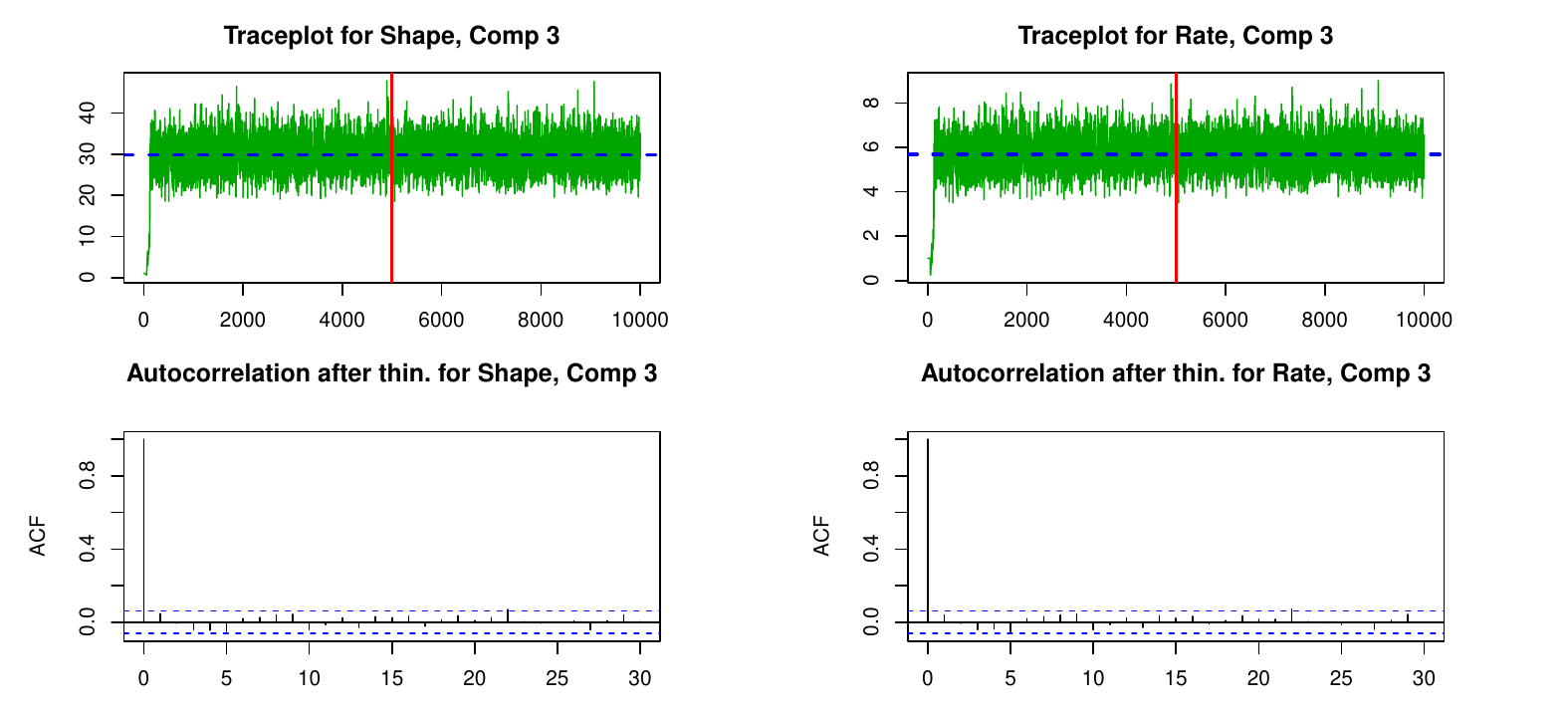}}\\
\caption{Results for the \code{asthma} data set showing the MCMC diagnostic plots, including traceplots and autocorrelation plots. }
\label{fig:asthma_diag}
\end{figure}

{We investigate the covariates' influence by examining the mixture probabilities from the last MCMC iteration. %, presented in Figure~\ref{fig:heatmap_asthma}.
An interesting discovery is that the mixture probabilities for both sexes, BMI levels, and severity levels are the same, indicating that the levels of these three covariates do not strongly influence the distributions of state durations.  
This matches the global test results in Figure~\ref{fig:global_asthma}.  
If the preceding state is state 1, which is optimal control for asthma, the state duration time is longer than that if the previous state is 2 or 3, as there is a lower weight in component 1 and higher weight in components 3 and 4. 
On the other hand,  if the preceding state is 3, which is unacceptable control, then the state duration time is much shorter, as the mixture probability in component 1 is much higher when the preceding state is state 3. 
We present some examples of the diagnostic plots for the \code{asthma} data set in Figure~\ref{fig:asthma_diag}.
}

\begin{example}
R> sm.asm$dur.mix.probs
$Severity 
       Mild-Moderate Severe
Comp 1          0.37   0.37
Comp 2          0.26   0.26
Comp 3          0.20   0.20
Comp 4          0.17   0.17

$BMI 
       BMI<25 BMI>=25
Comp 1   0.37    0.37
Comp 2   0.26    0.26
Comp 3   0.20    0.20
Comp 4   0.17    0.17

$Sex 
       Women  Men
Comp 1  0.37 0.37
Comp 2  0.26 0.26
Comp 3  0.20 0.20
Comp 4  0.17 0.17

$prev_state
          1    2    3
Comp 1 0.27 0.33 0.51
Comp 2 0.23 0.33 0.23
Comp 3 0.31 0.20 0.09
Comp 4 0.19 0.15 0.17

R> diag.BMRMM(res.fp2, cov.combs = list(c(1, 2, 1)), 
              transitions = list(c(1, 1)), components = c(3))
\end{example}

\section{Conclusion} \label{sec:end}

We presented the \CRANpkg{BMRMM} package which implements 
both Bayesian Markov mixed models (BMMM) for analyzing the state transitions 
and Bayesian Markov renewal mixed models (BMRMM) for additionally analyzing the duration times (being either state persistence times or inter-state intervals)
in a collection of categorical sequences, 
using flexible Dirichlet and gamma mixtures, respectively. 
The BMRMM takes into account fixed effects of the associated covariates as well as random effects of the associated individuals while
simultaneously selecting the significant covariates separately for the state transitions and the {duration times}. 
{The package includes a synthetic \code{foxp2} data set to demonstrate the data framework and function usages.} 
The package also provides a series of plotting functions for visualizing the results of the analyses, 
including various global and local hypotheses tests, MCMC diagnostics, etc. 
We are committed to maintaining and further developing the package in the future. 
{Future improvements to the package may include more options for the distribution types of transition probabilities and duration times beyond the currently available mixture Dirichlet and mixture gamma distributions, respectively.}

{\section{Acknowledgements} \label{sec:ack}
We thank two anonymous reviewers very much for their careful review of our work 
and their constructive comments that led to significant improvements to both the package and this paper.}

\bibliography{BMRMM_Package}

\address{Yutong Wu\\
  Department of Mechanical Engineering\\
  The University of Texas at Austin\\
  204 E Dean Keeton St C2200, Austin, TX 78712-1591\\
 United States\\
  ORCID: 0000-0001-7828-9981\\
  \email{yutong.wu@utexas.edu}}

\address{Abhra Sarkar\\
   Department of Statistics and Data Sciences\\
  The University of Texas at Austin\\
  2317 Speedway D9800, Austin, TX 78712-1823\\
  United States\\
  ORCID: 0000-0002-6924-8464\\
  \email{abhra.sarkar@utexas.edu}}